\documentclass[11pt]{article} 
\usepackage[T1]{fontenc}
\usepackage{libertine}
\usepackage[left=2cm,right=2cm,top=2cm,bottom=2cm]{geometry}
\usepackage[pdftex]{graphicx} 
\usepackage{amssymb} 
\usepackage{amsmath} 
\usepackage{amsthm}
\usepackage{wrapfig} 
\usepackage{calc} 
\usepackage{footmisc} 
\usepackage[table, svgnames, dvipsnames]{xcolor}
\usepackage{makecell, cellspace, caption}
\usepackage{subcaption}
\usepackage{lineno}
\usepackage{mathtools}
\usepackage{url}
\usepackage{hyperref}
\usepackage{appendix}
\hypersetup{colorlinks,linkcolor={},citecolor={blue},urlcolor={red}}

\usepackage[version=4]{mhchem}

\usepackage{booktabs}
\usepackage[square,sort,comma,numbers]{natbib}

\usepackage{epigraph}
\usepackage{etoolbox}
\usepackage{authblk}

\makeatletter

\patchcmd{\maketitle}{\@fnsymbol}{\@alph}{}{}
\makeatother

\title{On the regularity of human mobility patterns at times of a pandemic}

\author[a,b]{Fabio Vanni}
\author[b,c]{David Lambert}

\affil[a]{Sciences Po, OFCE , France}
\affil[b]{Center for Nonlinear Sciences, University of North Texas, USA}
\affil[c]{Department of Mathematics, University of North Texas, USA }

\date{}

\begin{document}
\maketitle

\begin{abstract}
The study of human mobility patterns is a crucially important research field for its impact on several socio-economic aspects and, in particular, the measure of regularity patters of human mobility can provide  a across-the-board view of many social distancing variables in epidemics such as: human movement trends, physical interpersonal distances and population density. We will show that the notion of information entropy is also strongly related to demographic and economic trends by the use and  analysis of real-time data. 
In the present research paper we address three different problems. First, we provide an evidence-based analytical approach which relates the human mobility patterns, social distancing attitudes and population density, with entropic measures which depict for erraticity of human contact behaviors. Second,  we investigate the correlations between the aggregated
mobility and entropic measures versus  five external economic indicators.
Finally, we show how entropic measures represents a useful tool for testing the limitations of typical assumptions  in epidemiological and mobility models.  
\end{abstract}


\section{Introduction}
A surge of interest has been noted in the use of mobility data from mobile phones to monitor physical distancing and
model the spread of severe acute respiratory syndrome coronavirus 2, the virus that causes COVID-19. 
The study of mobility can provide epidemiological relevant estimates about  the extent to which people are sheltering in place and generally moving differently from  an usual baseline.  Mobility data also give useful information about travel patterns to help better understand the effect of travel restrictions and the risk of importation from other locations.
Human mobility and, subsequently, social interactions can be seen both random and regular, showing an intrinsic complexity which can be captured by a physics or information theory based measure for the fluctuations as the entropy which can help in understanding  regularity of individuals' visiting patterns.  In particular, there is substantial value in having a metric, that captures degree and patterns of employees' and consumers' mobility since the commercial value of understanding motion is substantial. Magnitude and patterns of mobility turn out to be powerful predictors of human behavior, specifically behaviors that businesses seek to drive. 
In this work we study features of individual mobility that can be related to entropy measures and so capturing the relationship within human mobility patterns in the context of socio-economic Covid-19 pandemic crisis. Cellular phone data offers an ubiquitous opportunity for scientists to observe how people re-act, move and respond to external influences like a epidemiological shock. Such data can be used as measurement of interactions of individual persons in a social context. Studying the effects of social distancing measures on the evolution of the epidemic diffusion, as in the case of the Covid-19 pandemic, has been undergone into controversial academic and political positions. There are  many factors which can influence the effectiveness of spontaneous or mandatory behavioral social responses which are difficult to be considered separately as definitive evidences against or in favor about the effectiveness of  non-pharmaceutical interventions. They need to be considered together to describe a composite effect on contagious transmission. Among the social-related factors which have an impact on the spread of airborne diseases like the Covid-19 case, we find human mobility, inter-personal distance and population density and usually in epidemiological modeling, such composite effects  are essential summarized as interaction rate among individuals.  Under a collisional modeling of human contacts we can point out a useful statistical and analytical relation between entropy measures and some of the social distancing variables where mobility plays a crucial, but not unique, role. We then show how an entropy-based approach can provide a better insight into the analysis of socio-economic indicators.
In particular, we start from  mobile phone data at US state level, using two mobility measures: radius of gyration, the characteristic distance traveled by an individual, and mobility entropy, the diversification of movements over her locations. 
Human economic behavior is affected by human geography since constraints on mobility determine if people can go to work and what they can buy and, in turn,  shopping choices drives people movement. So, describing consumer patterns can be important both for modeling the dynamics of a market, and for discovering predictability of future behavior at the individual level.
Economic models of consumption incorporate constraint and choice to varying degrees. 
We compare the aggregated mobility measures with some socio-economic indicators measuring as employding, firm revenues, energy demand and the coincidence index. Performing a correlation analysis, we observe a relationship of such economic indicators against both mobility and entropy-based variables. Such correlations shows a varying intensity according to the drivers of each economic indicator.  In fact, other than the mobility, very important demographic variables are repredented by population density and personal space of individuals (i.e. interpersonal distances). However,  those variables are not so unambiguously traceable as real-time data since they presents some intrinsic difficulties in collecting proper data to be used directly in our analysis. However, we prove that they are embedded into the visiting patterns encapsulated in the mobility entropy.
As final result, our analytical analysis is grounded on a kinetic collisional model where many mobility and demographic variables are used. By the way,  such theoretical framework rests on some fundamental assumptions, the most important of which consists in the hypothesis of random and uncorrelated contacts. By using the concept of information entropy we can check  how much evidences from real data deviate from such erratic assumption of social movements, so evaluating  in what extent aggregated mean field model works are reliable in describing real world dynamics.

\section{Literature review}
The study of human mobility and its regularity is encoded in an interdisciplinary field that aims to understand the intrinsic properties of human  movements as well as the mechanisms behind the observed patterns. The concept of human mobility  encompasses various dimensions of human travel at both individual and group levels. 
 In fact, there is a very important emergent literature on an interdisciplinary  research about regularity of human mobility. Some theoretical studies as summarized by \citet{barbosa2018human,gallotti2013entropic,osgood2016theoretical} who discussed  the study, understanding, and modeling of human mobility. In particular the study of human mobility has received considerable attention in terms of indicators based on the concept of entropy  directly applied to real world mobility flows as discussed by \citet{song2010limits,lu2013approaching,kulkarni2019examining}.
Mobility is usually associated as the average distance traveled by individual or other alternative definition, and it ,in fact, represents an important proxy measure of social distancing. As consequence, mobility contraction has represented one of the crucial social distancing measures for restricting population movement thus reducing the number of contacts and consequently reducing the transmission of SARS-CoV-2, see \citet{gatalo2020associations,badr2020association,vanni2020epidemic,bonaccorsi2020economic,cintia2020relationship,nouvellet2021reduction}.

Despite the fact that human mobility is seen as a crucial property of a population, it is actually only one ingredient of what in general one could see as human movement patterns, for example, the role of population density. Recent studies has proven controversial results on the role of population density on the spreading and/or death rate of the population density \citet{rader2020crowding,carozzi2020urban,Gerritse2020covid,wong2020spreading,bhadra2021impact}. 
From one side, the individuals living in high population density areas have a higher probability to come into close contact with others and consequently any contagious disease is expected to spread rapidly in dense areas. So, in principle,  population density may affect case and death rates since people have contacts in closer proximity to each other so that the virus SarsCov2 is transmitted through droplets proximity to other individuals is one of the risk factors. As consequences, in places where it is more difficult to practice social distancing, like densely populated urban areas, one expect to observe higher incidence. On the othr hand,  there are some states where rural communities are actually the ones experiencing disproportionate infection rates often because of local outbreaks and exacerbating underlying conditions.  
In fact, despite the direction of relationship between epidemic diffusion and population density is ambiguous, there are  other mediating factors that might affect the behavioral responses to the pandemic, which can itself affect the spread and severity of the outbreak.

Many researchers \citet{chu2020physical,gokmen2020national,cartaud2020wearing,welsch2020physical,de2021covid,xu2012does,kishore2020measuring}, have investigated the effects of physical distance on the Covid-19 disease transmission, since the risk for infection is highly dependent on distance to the individual infected and the type of face mask and eye protection worn.  It is well assessed how there exists a significant and negative (opposite) relation between national interpersonal distance preferences and COVID-19 spread rate. The inter-personal distance  is the  object of proxemics study of human use of space and the effects that population density has on behaviour, communication, and social interaction.

On the other hand, the rise of widespread computing techniques provides an up-to-date and accurate way to detect human movements at various temporal and spatial scales.  This allows for acquisition of new knowledge about important aspects of human mobility patterns, so that the resulting location datasets can be then used to study and model user mobility behaviors. \citet{cintia2020relationship,pappalardo2016analytical,luca2020deep,bonato2020mobile}. Real-time user locations are typically collected using the global positioning system (GPS), call detail record logs (CDR) and wireless-LAN (WLAN). In fact, recording human activities can yield high-fidelity proxies of socio-economic development and well-being. However mobility data have their biases and limitations, for example they can be more representative of a younger and more affluent population, but at the same time another data stream could under represent those living in rural areas.

The analysis of human regularity of mobility  during the outbreak of an epidemic  is crucially important in social sciences and economics because of the strong interaction between aggregate demand and the dynamics of an epidemic so that the epidemic trend affect the consumer's behavior and, in turn, the demand trend affects the amount of physical contacts across individuals and hence the epidemic incidence \citet{jiang2021impacts,heroy2021covid,kramer2020potential}.  The rising aggregate demand increases the contact rate and therefore the exposure to a virus infection.  On the other hand, rising infection lowers aggregate demand because of  reduced  household  spending. On the contrary,  higher epidemic incidence depresses aggregate demand, which lowers the contact rate and reduces infections. 
Furthermore, some recent studies have shown that human movement patterns are strongly associated with regional socioeconomic indicators \citet{brodeur2020literature,papageorge2020socio,basurto2020economic,avery2020economist}. In particular, direct  and  indirect effects of lockdown measure has trigger economical sequential  adjustment  process  in  response  to  shocks  to productive capacity  (supply  shocks) and/or  to final  demand (demand  shocks). There already exists a rapidly growing literature examining the economic effects of Covid-19, with  joint  analyses  of  economic  and  epidemiological  dynamics as well as of the possible interconnections between supply and demand shocks originating in the pandemic \citet{eichenbaum2020macroeconomics,guerrieri2020macroeconomic,reissl2021assessing,flaschel2021pandemics}.  The epidemic has had a simultaneously impacting both the supply and demand sides of the affected economies, mostly caused by mobility contraction seen as movements reduction, physical distancing increase and confinement policy of stay-at-home orders. Such mandate measures can bee seen as shocks affecting the sectoral availability of labor, constrain the output that can be allocated to final demand  and  to  other  sectors,  affecting  demand  for  non-labor  inputs  and,  possibly,  creating bottlenecks in the production of downstream sectors. 
The economic and social shock presented by the Covid-19 pandemic has reshaped  perceptions of individuals and organizations about work and occupations, resulting in changes of occupational supply and demand and changes  in occupational perspectives on working from home. Moreover, employment rate is strongly related with  aggregated measures of consumer spending \citet{krumme2013predictability,dynarski1987consumption,amadeo2019consumer} especially during times of crisis.

As a further consequence, the mobility reduction, physical isolation, economic difficulties (as unemployment) have an important effect on the individual health and it can represents in its turn a social and economical problems as well, \citet{brenner2020acceleration,auriemma2020covid,newbold2020effects,welsch2020interpersonal,kroczek2020interpersonal} where there exists a relationship between the extent of physical distancing and lost income in society. So if from one side social distancing measures can safeguard public health, they also can profoundly impact the economy and may have important indirect effects on the society, posing also serious challenge to behavioral norms. For example, a close interpersonal distance  increases emotional responses during interaction and has been related to avoidance behavior in social anxiety.
Mobility data are also crucial to  understand climate change and migration patterns induced by climate extremes in order to understand the long-and short-term effects of climate change on vulnerable populations, \citet{gioli2016human,wang2016patterns}.


\section{Model and Results}
Entropy captures the degree of predictability characterizing a time series, and it was originally introduced to explain the inclination of intensity of heat, pressure, and density to gradually disappear over time. Entropy is a significant, widely used and above all successful measure for quantifying in-homogeneity, impurity, complexity and uncertainty or unpredictability.
We will focus the attention on Entropy variable which allows to characterize  many aspects of population dynamics such as human mobility, population density, inter-personal distance (proxemics) and minimal movement trend of individuals. In this terms entropy can also be seen as measure of attractiveness and socioeconomic complexity.

\begin{table}[!ht]
	\centering\captionsetup{justification = centering}
	\caption{Data Repositories at state US level }
	\label{tab_data}
	{
		\setlength\arrayrulewidth{.001pt}
		\begin{tabular}{l}
	 \toprule 
			\rowcolor{black!5}
			\makecell{\textsl{Mobility Data}}      \\ 
			{{Camber Systems Social Distancing Reporter \cite{camber}}}- \textsl{Entropy \& RoG} (\textit{daily}) \\
			 {SafeGraph  Data Consortium \cite{safegraph}} - \textsl{Visited locations \& Distance traveled} (\textit{daily}) \\
			 U.S. Census Bureau \cite{census2011} - \textsl{Populaton density} (\textit{annual})\\
			  \hline
			\rowcolor{black!5}\midrule
			\makecell{\textsl{Economic Data}}      \\ 
			Economic Tracker \cite{economictracker} - \textsl{Employment, Consumer Spending \& Firms Revenue} (\textit{daily}) \\ 
			U.S. Energy Information Administration \cite{iea} - \textsl{Energy Demand and Production} (\textit{daily}) \\ 
		Federal Reserve Bank of Philadelphia \cite{federalphi} - \textsl{Coincidence index} (\textit{monthly})\\
			\bottomrule		
		\end{tabular} 
	}
\end{table}

The importance physical distancing practices is directly related with the epidemiological trends since mobility and population gatherings are essential features for contagion diffusion, and as expressed in , one can write the reproduction number and the economic damage of the effects of the mitigation policy to contrast the epidemic as:
\begin{align}
R_{t} &= f(s_t ,\mathcal{B}_t)  &  \text{ Epidemic repreduction number} \label{eq_intro}\\
{D}_t&=f(\omega_t,\mathcal{L}_t) & \text{ Economic damage function} \label{eq_intro2}
\end{align}
As regard with the epidemiological side, and following \citet{vanni2020epidemic, palatella2021phenomenological}, the reproduction number is a function of  $s_t$ which is the share of susceptible population (not immunized individuals), $R_{t_0}$ is the reproduction number at the beginning of the observed period, and $\mathcal{B}_t$ is the transmission rate function which depends on variables which describe the interaction frequency of contacts and the rate of becoming infected after a contact. Furthermore it also depends on the infectious age of individuals in the contagion process. On the other side, the economic damage function $D_t$ is a not well specified function which would account for the economic impact of the pandemic as discussed \citet{bellomo2020,reissl2021assessing}, where, specifically, the damage (or loss) function depends on the intensity of the lockdown $\mathcal{L}_t$ (social distancing restrictions) which is inversely related with the interaction patterns in society. We put the basis to connect the transmission rate $\mathcal{B}_t$ to the lockdown intensity $\mathcal{L}_t$ through a dashboard of mobility variables,  since a reduction in social movements have impacts both on the epidemiological side and the economical one. A possible candidate for a such connection can be found in the entropy-based metrics. Finally, $\omega_t$ represents the specificity of the particular  economic indicator under investigation and it captures the sensitiveness of that particular economic indicator to mobility and  social distancing changes.

In the present research paper, we express the concept about how the transmission rate of contacts $\mathcal{B}_t$ can be expressed in terms of social distancing measure ans consequently in terms of entropy.  On its turn,  we shed lights on the crucial role of information entropy of mobility patterns in relation between  lockdown and economic activity.
We make use of analytical tools supported by statistical evidences which relies on the use of large datasets for both mobility and economic data, as reported in Table \ref{tab_data}.

\subsection{Mobility pattern estimates }
Thus the entropy $S(X)$ is equal to the amount of information learnt on an average from one instance of the random
variable $X$. It is important to highlight that, the entropy does not depend on the value that the random variable takes, but only on the probability distribution $p(x )$.  The probabilities of different values can
be leveraged to reduce the number of bits needed to represent the data if and only if the variable has non-uniform distribution. Thus, entropy can also be defined as the measure of compressibility of the data, or a measure that defines the predictability of a single random variable. Lower entropy therefore generally signifies higher predictability, meaning that an individual's time spent at their significant  locations  is  highly  predictable.  Conversely,  high  entropy suggests that predicting an individual's location   is   difficult.   Therefore, the lowest    entropy  would  be  achieved  by  a  user  who  spends  the  exact  same  amount  of  time  in  the  same  places  in  every  time window. You can stay in a small area but be moving unpredictably or traverse a large area very predictably. Someone just staying at home, or just going to work from home, is extremely predictable (though they may be traveling far from home).  The entropy metric tracks how erratic or how predictable a person's movement is. Entropy will be higher when someone moves around with less consistency, and it will be lower when they move around with more predictability.

In the study of human mobility, random entropy, measures the uncertainty of an individual's next location assuming that this individual's movement is completely random among $L$ possible locations, and is calculated as:
\begin{equation}
S_{\text{rand}}=\langle \log L_i \rangle_N
\end{equation}
This, differently,  captures the predictability of each user by assuming that the person's whereabouts are uniformly distributed among $L_i$ distinct locations.

\begin{figure}[!ht]
	\centering
	\includegraphics[angle=0,origin=c,width=0.45\linewidth]{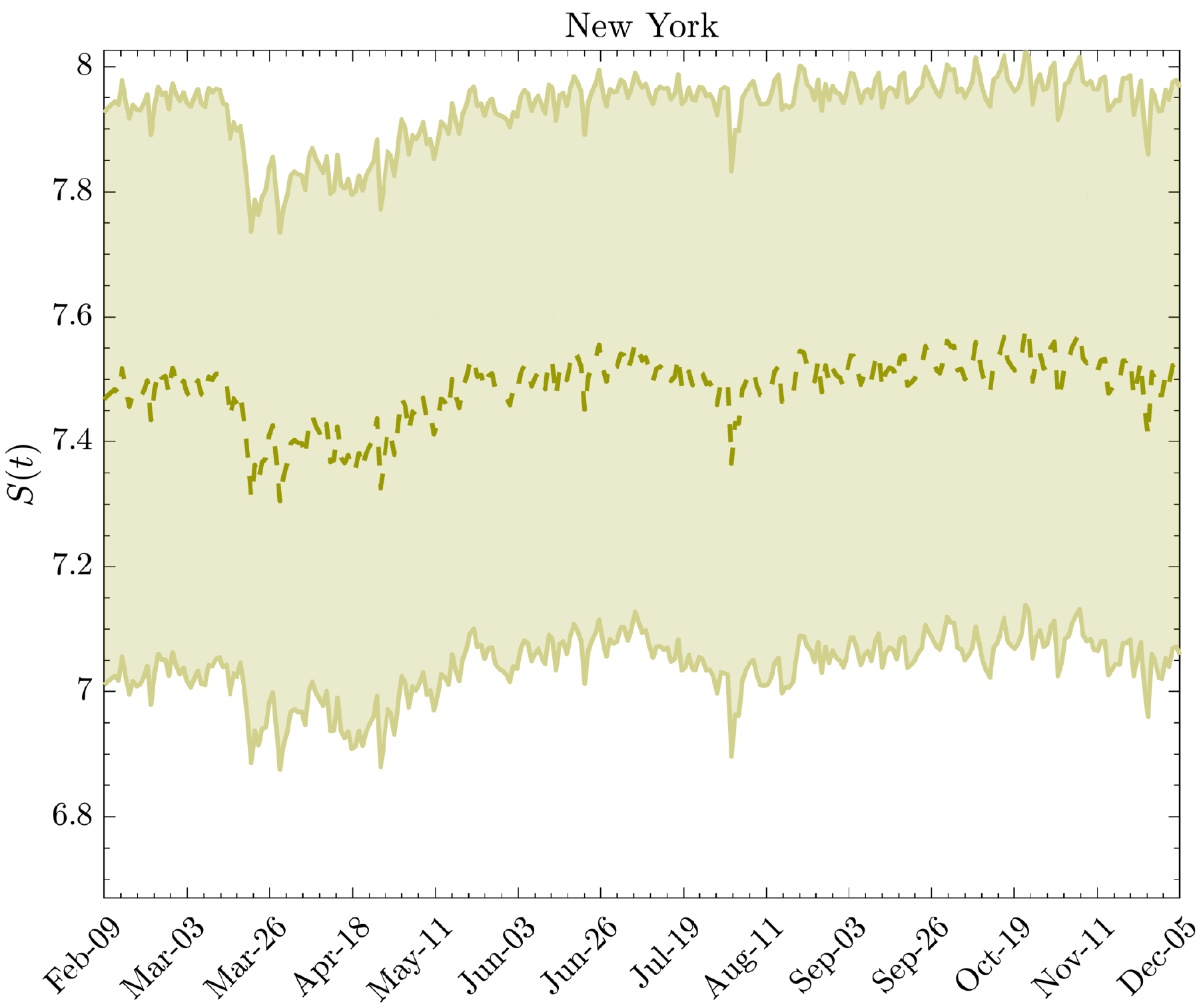}
	\includegraphics[angle=0,origin=c,width=0.45\linewidth]{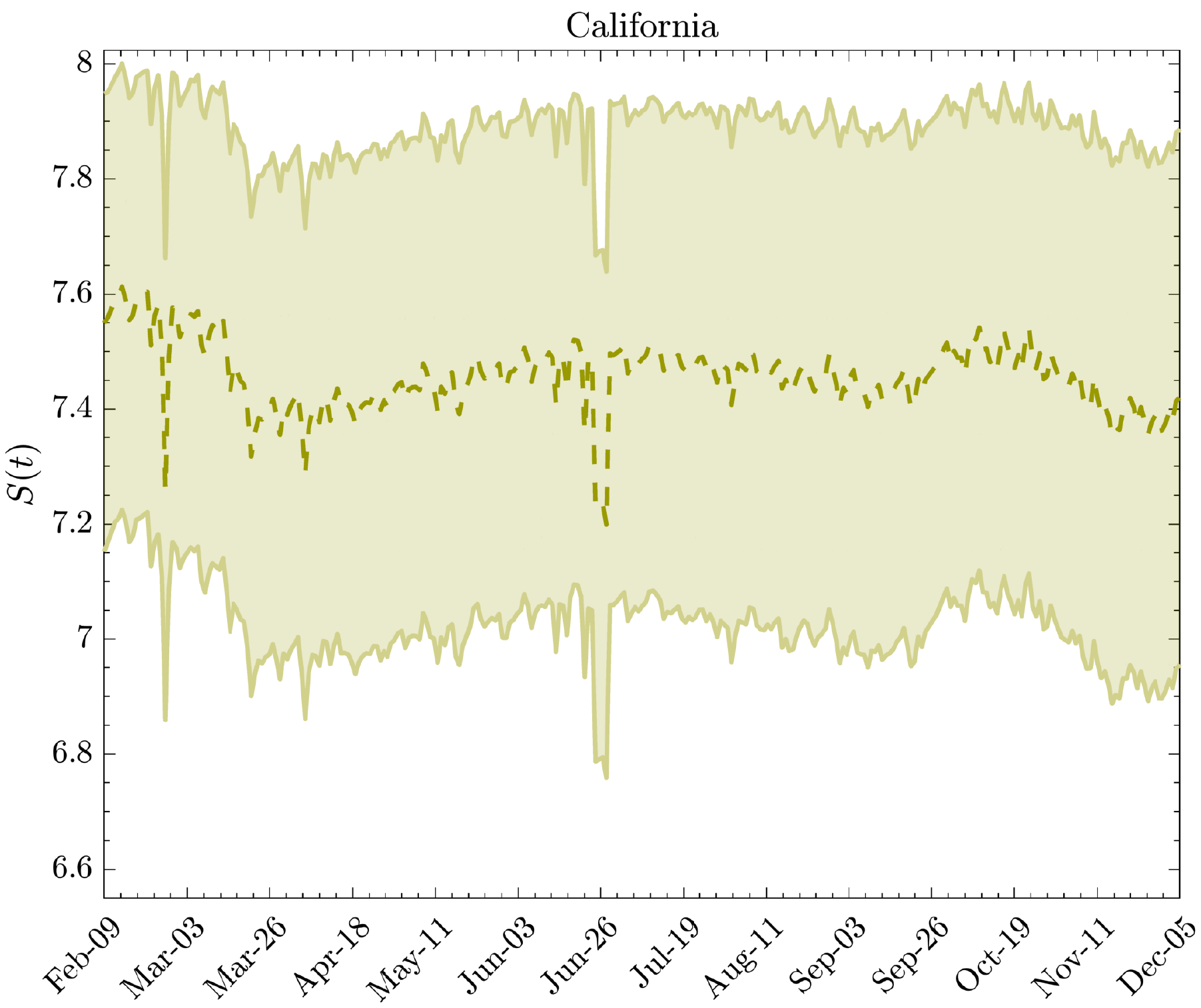}
	\includegraphics[angle=0,origin=c,width=0.45\linewidth]{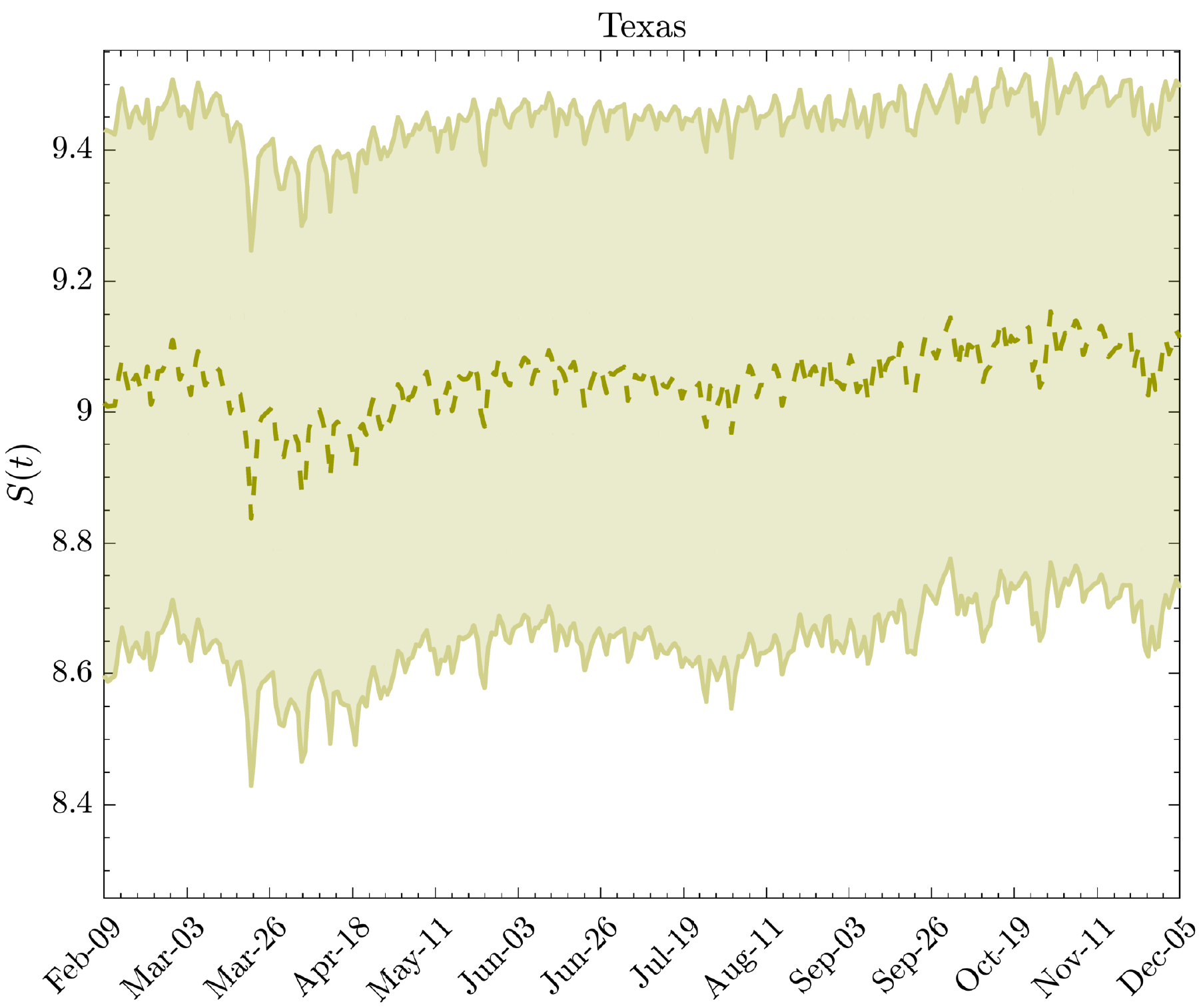}
	\includegraphics[angle=0,origin=c,width=0.45\linewidth]{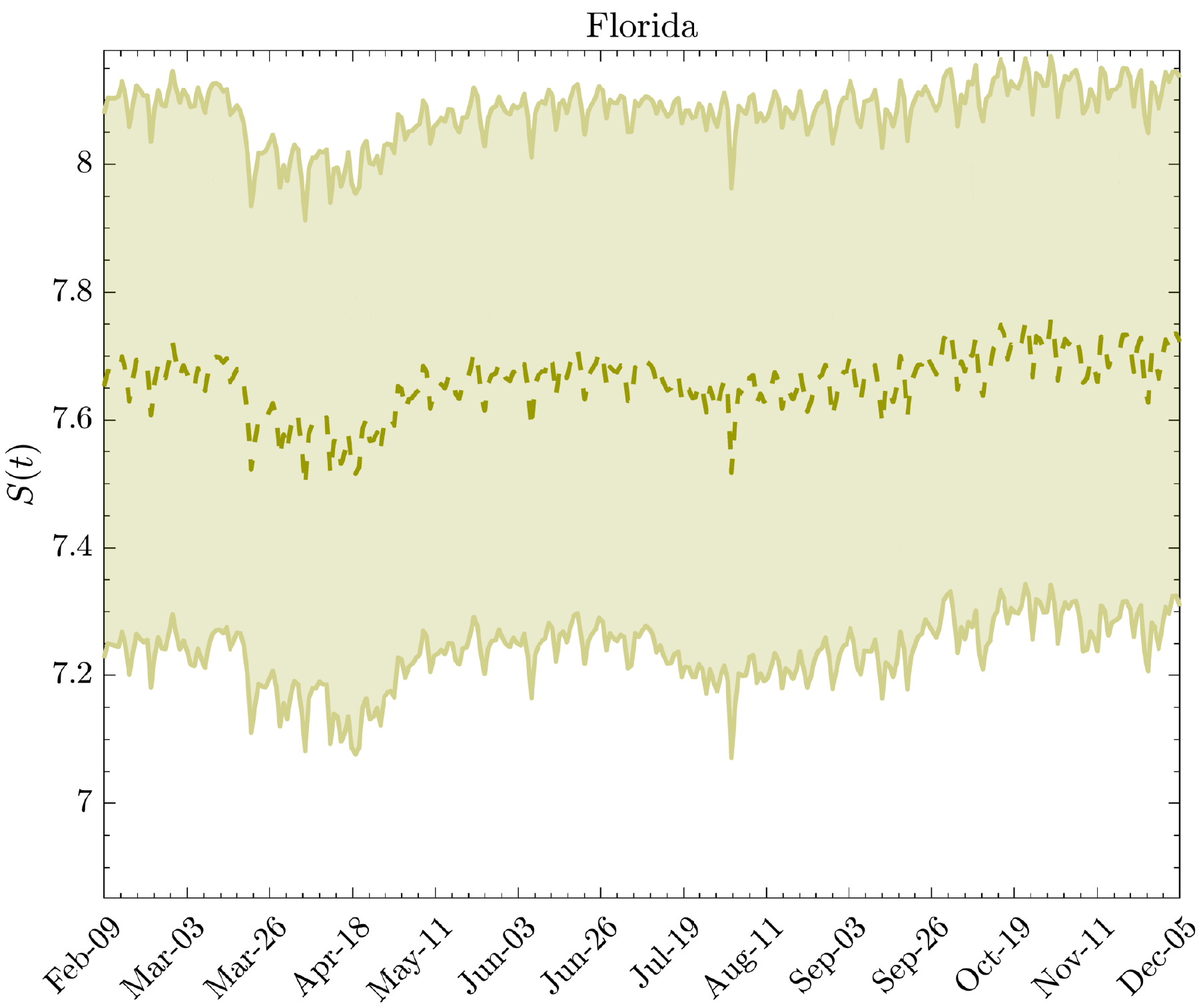}
	\caption{Median Uncorrelated Shannon entropy of individuals' movements, data courtesy of Camber Systems \citet{camber}.}	
	\label{fig_entropy}  
\end{figure}

If the individual's movement among N possible locations follows a certain probability distribution,  the entropy  of this process is then defined as:
\begin{equation}
S_\mathbb{U}=-\left\langle \sum_{k=1}^{L_i}p_k\log p_k \right\rangle_N \nonumber
\end{equation}
where $L_i$ is the number of distinct locations by each of the individuals, $i=1\ldots N$, and where $p_k$ is the frequency of the user's visit to their $k$-th location ($k$ is the index of all locations that the user visits). 
 In particular, we plot the uncorrelated Shannon Entropy measure for some US states in Fig.\ref{fig_entropy}. Shannon entropy is high when an individual performs many different trips from a variety of origins and destinations; it is low when he performs a small number of recurring trips. The uncorrelated Shannon entropy  takes into account the number of different locations visited as well as the proportion of time is spent at each location, so decreasing the uncertainty of the trajectory respect to the definition of the random entropy.  Let us notice that it is always true that $S_\mathbb{U}\leq S_{\text{rand}}$ where the equality holds  when the process is completely random.

Additionally, we could also considered movement patterns where  one considers the frequency of the visited locations and the order as well. However, we have used data provided by Camber Systems database \citet{camber} and partially from SafeGraph \cite{safegraph}, from which  it is possible to report both the random and the temporal uncorrelated Shannon entropy in meters of GPS persons' devices. However it is not possibile at this stage to recover individual patterns for the periods of Covid-19 pandemic, which could allow us to study the change in time-correlated paths of individuals, see \ref{app_real} .

Analytically we can compute the entropy from a collisional model where individuals move in a region much larger than the their interaction regions and after some calculations, see \ref{app_config}, it is possible to write the configurational entropy in terms of information entropy for our equiprobable system as: 
\begin{equation}\label{eq_enstropyconf}
E[S_{rand}] \approx 1+\log\left(\frac{1}{\delta} -\frac{1}{\delta_1}\right) \qquad \text{ for } \;\;\delta \ll \delta_1 
\end{equation} 
where $\delta=N/A$ is the (effective) population density and  $\delta_1$ is defined as the minimal area occupied by a single individual (a space where no other individual is present but you) since $\delta_1=1/\pi r^2 $ where $r$ is the interaction radius which is inversely proportional to the typical interpersonal distance \citet{sorokowska2017preferred}. More precisely, it corresponds to resolution of GPS on revealing such area. Moreover,  social distancing measures can modify the interaction radius and so, consequently, the value of the entropy. More importantly, in our representation, the variable $r$ can be affected by other factors as cultural, structural, climatic,environmental, etc.  However, the  configurational entropy is just an approximation of the random entropy since we cannot know the exact number of urban population density as fully discussed in \ref{app_pop}. As a consequence we cannot directly use the population density reported by US Census Bureau \citet{census2011}, which are calculated in terms of physical and geographic region and not the effective area where people interact. As possible  improvement is represented by the notion of population weighted density  that is the population density that the average person experiences, \citet{ottensmann2018population}, measuring the density at which the average person lives. 
Furthermore, the configurational entropy should be discounted for the people staying in lockdown (which does not move so they do not participate to the collisional movements) so capturing  the change in the interaction radius $r$ during the evolution of the epidemic. However, defining the  population density $\delta$ is actually a difficult task since we should map the number of individuals in an effective region of interaction. We do that by using the number of active devices reported by Camber System database.

Despite that, we know that in the limit of validity of our collisional model assumption  we know that $S_{\text{rand}}\ge S_{\mathbb{U}}$ where the equality holds if the probabilities $p_k$ are equally probable  (perfect disorder limit)\footnote{Precisely, $S_{\mathbb{U}}$ is a better measure of the entropy respect to $S_{\text{rand}}$, as it considers  the mobility spatial pattern revealing the uneven frequencies of the visits.  They are equals only if  all of the $L_i$ distinct visited places, $k$, were occupied with uniform probability $p_i(k) = 1/L_i$ by the individual $i$. However all those entropy measures  still ignore possible temporal patterns  which can be due to the relative position of the individuals, which is taken into account by the information entropies like real and conditional entropies, resulting in  $S_{\mathbb{H}}\leq S_{\mathbb{U}} $.}.

\begin{figure}[!ht]
	\centering
	\includegraphics[angle=0,origin=c,width=0.45\linewidth]{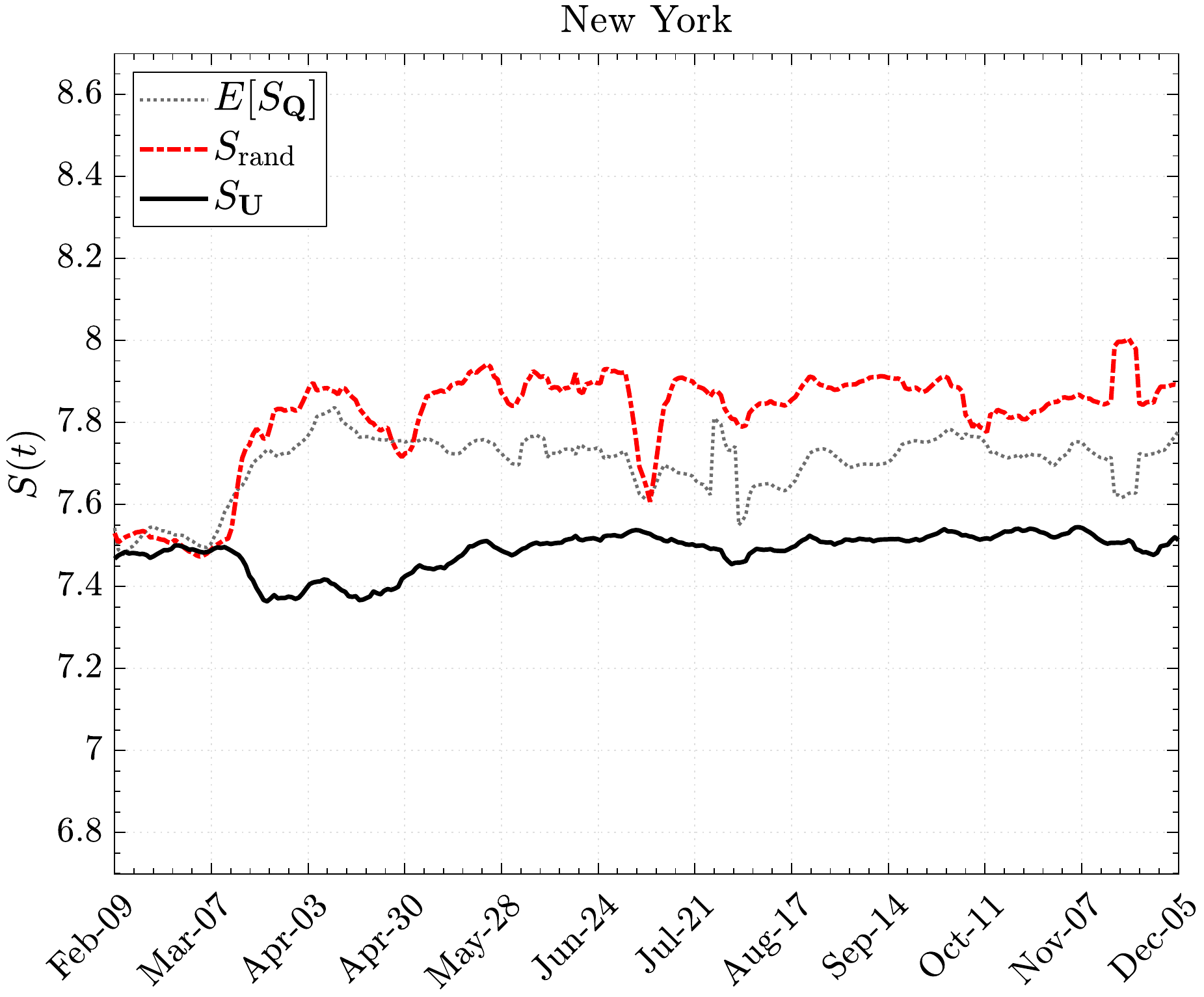}
	\includegraphics[angle=0,origin=c,width=0.45\linewidth]{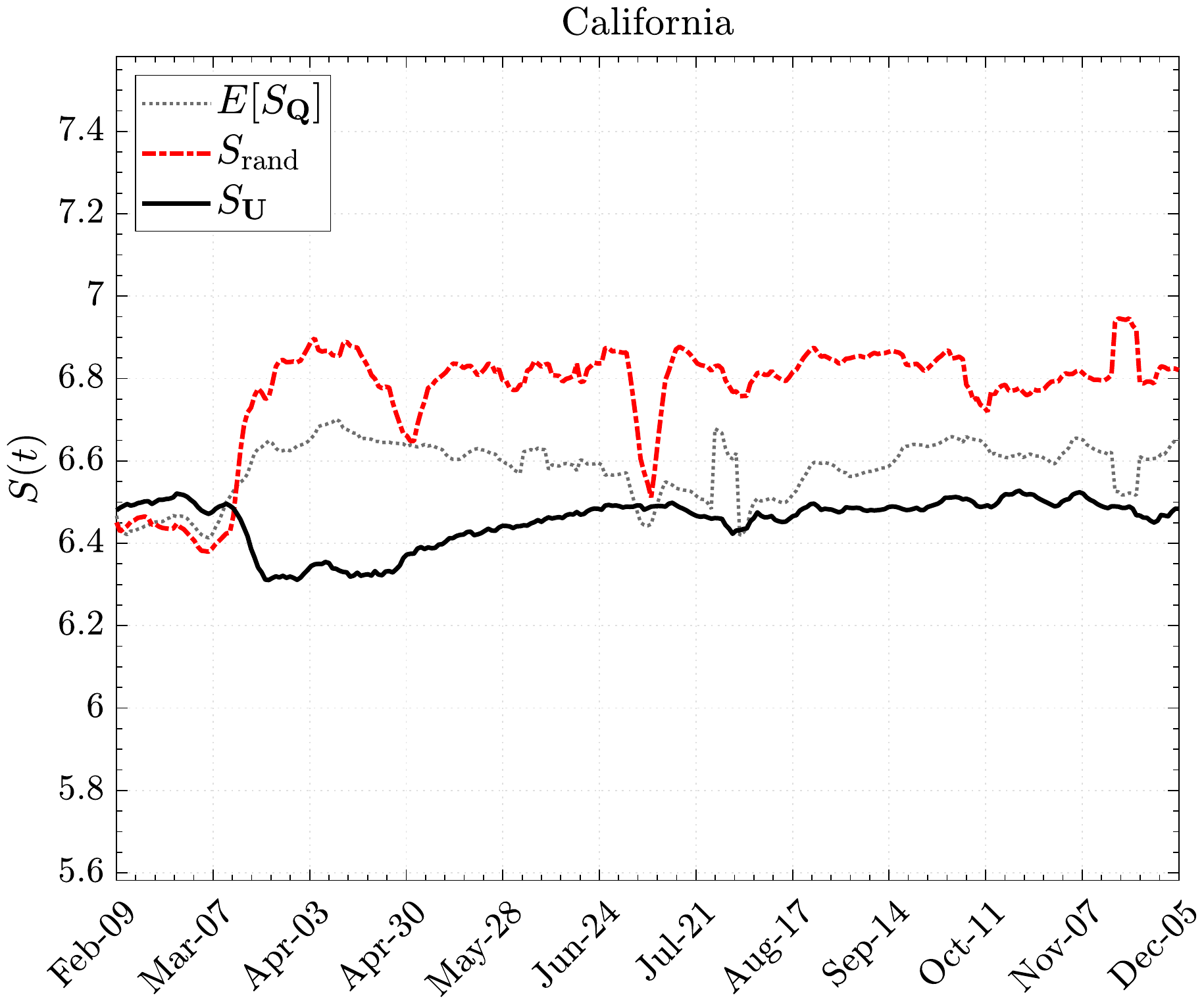}
	\includegraphics[angle=0,origin=c,width=0.45\linewidth]{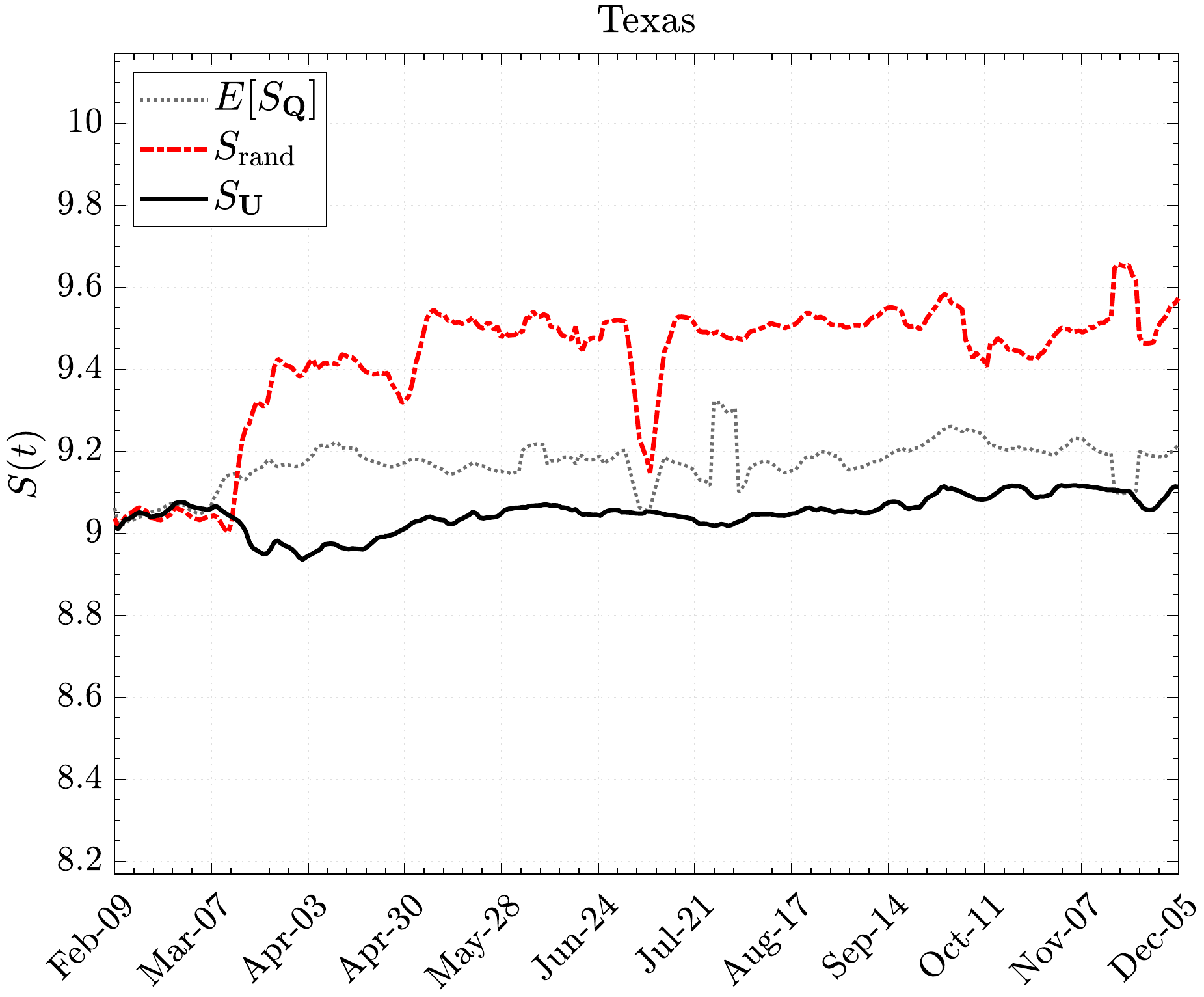}
	\includegraphics[angle=0,origin=c,width=0.45\linewidth]{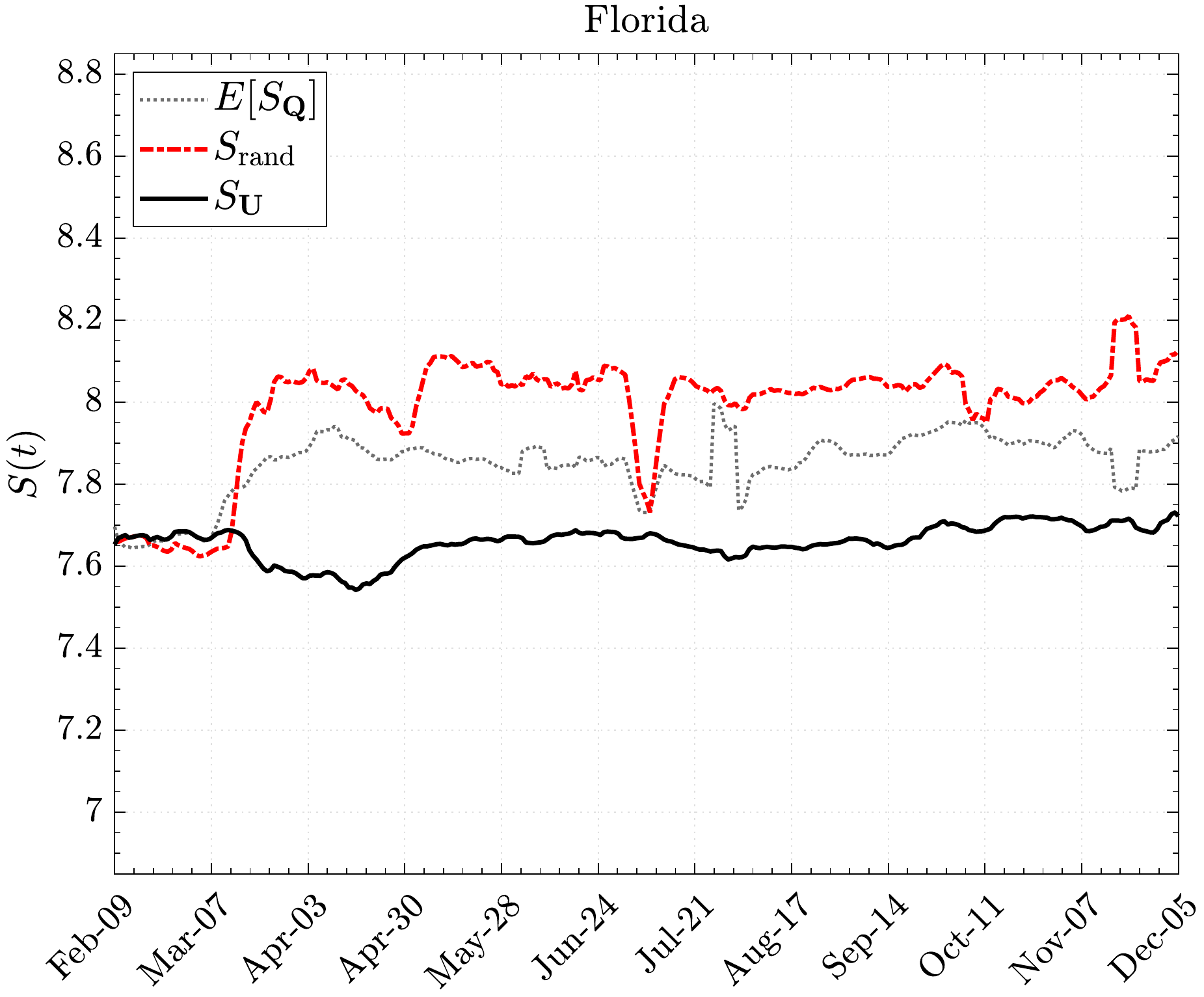}
	\caption{Comparison between the various implementation of Entropy of individuals' movements, data courtesy of Camber Systems \citet{camber}  and U.S. Census Bureau \citet{census2011} for population data.}	
	\label{fig_entropy3}  
\end{figure}

 For example we show in Fig.\ref{fig_entropy3}  all those entropic measures in some USA states. Despite the fact that such entropy measures looks similar in their mean value, there is significant difference in the trend behavior between the random and the uncorrelated Shannon entropies. In all the cases reported, we notice a change after the beginning of March 2020: the random entropy has increased meanwhile the uncorrelated one exhibits an opposite trend. A possible interpretation is that, after the Covid-19 outbreak individuals increased the number of stops (visited more distinct locations), but at the same time they have moved less uniformly  among differences locations they have stop by. This possibly reflects the fact that people have kept moving but the spend more time in important places like essential workplaces, and less time in visiting non-necessary locations (bars, restaurants etc...). By the way, during the year the uncorrelated Shannon entropy have recovered its original value, otherwise the random Shannon entropy has maintained is new mean value, which according to our configurational interpretation  model the density of active individuals has diminished.  

The Shannon entropy $S(k)$ is equal to the amount of information learned on an average from one instance of the random
variable $k$, but it does not depend on the value that the random variable takes, but only on the probability distribution $p(k)$. The probabilities of different values can be leveraged to reduce the number of bits needed to represent the data if and only if the variable has non-uniform distribution. Thus, entropy can also be defined as the measure of compressibility of the data, or a measure that defines the predictability of a single random variable. Lower entropy therefore
generally signifies higher predictability
\begin{equation}\label{eq_entropymobi}
E[S_{\mathbb{U}}] \approx \log\frac{\mu}{\mu_1} +\log\frac{\delta_1}{\delta}
\end{equation}
where $\mu$ is the aggregate mean mobility and $\delta$ is the population density of people actively participating in the social interaction. Then, $\mu_1$ is the smallest detectable mobility of a single individual (namely device), and the $\delta_1$ is the personal space of an individual which is considered as the average social area for each individual.

\begin{figure}[!ht]
	\centering
	\includegraphics[angle=0,origin=c,width=0.45\linewidth]{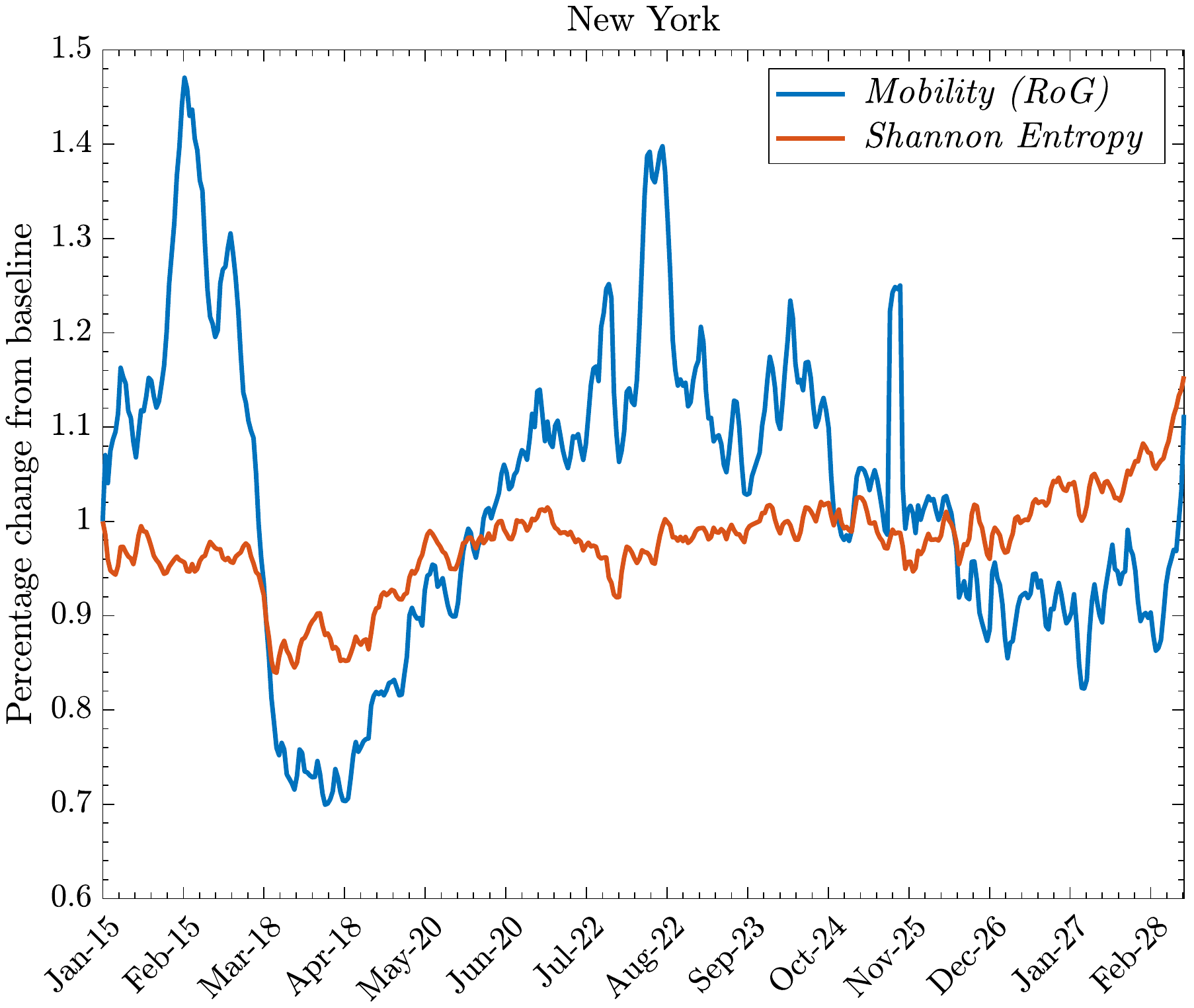}
	\includegraphics[angle=0,origin=c,width=0.45\linewidth]{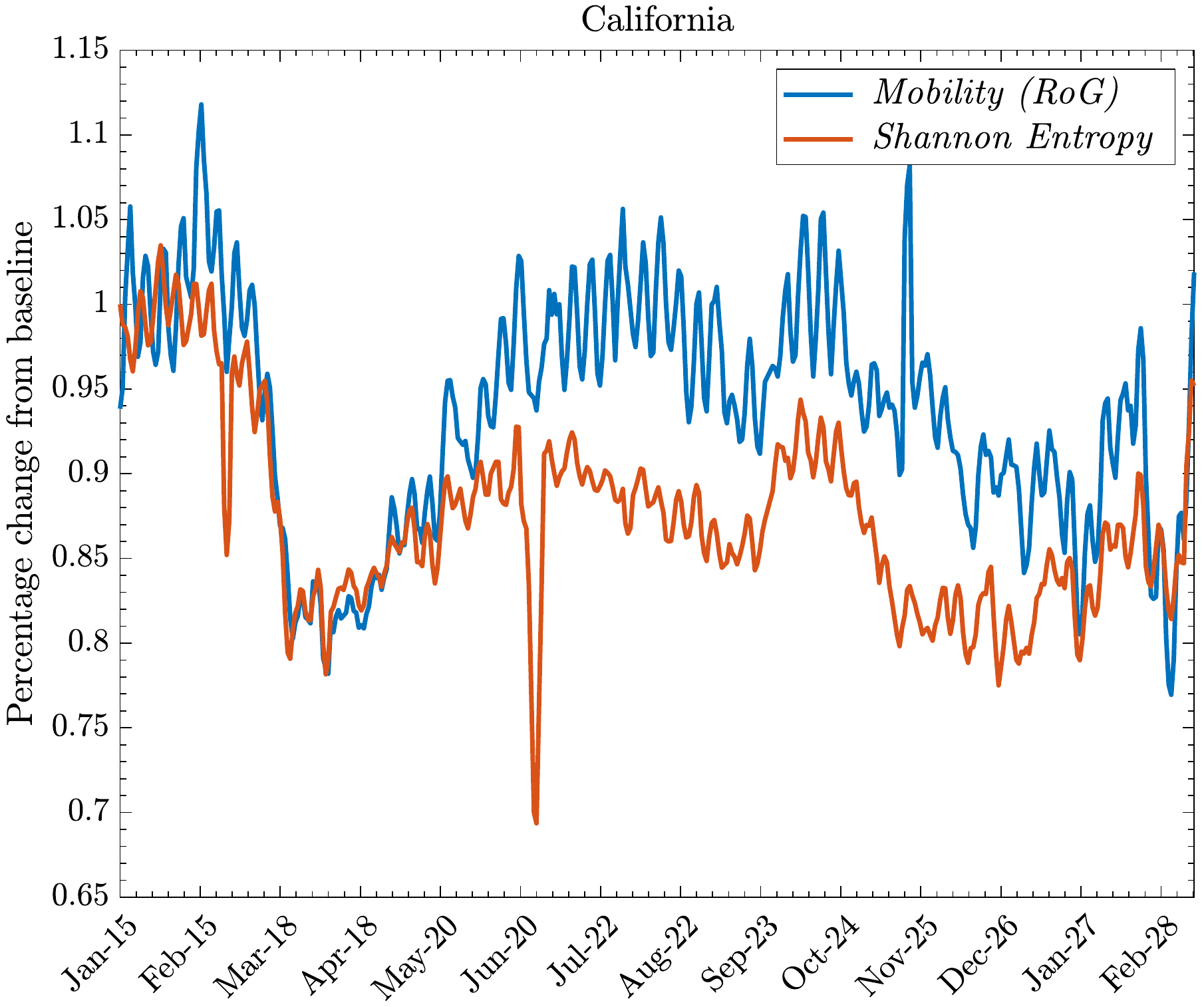}
	\includegraphics[angle=0,origin=c,width=0.45\linewidth]{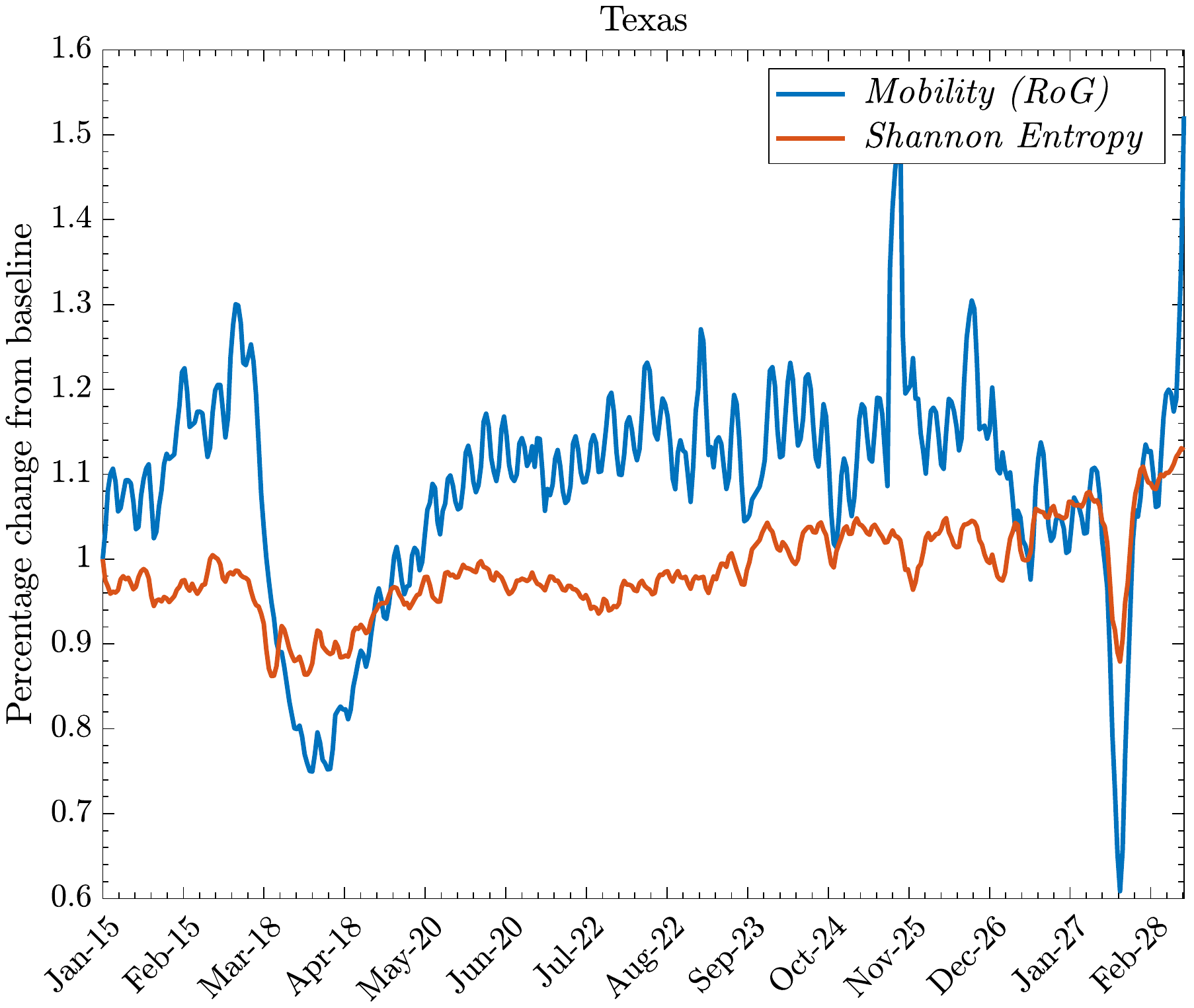}
	\includegraphics[angle=0,origin=c,width=0.45\linewidth]{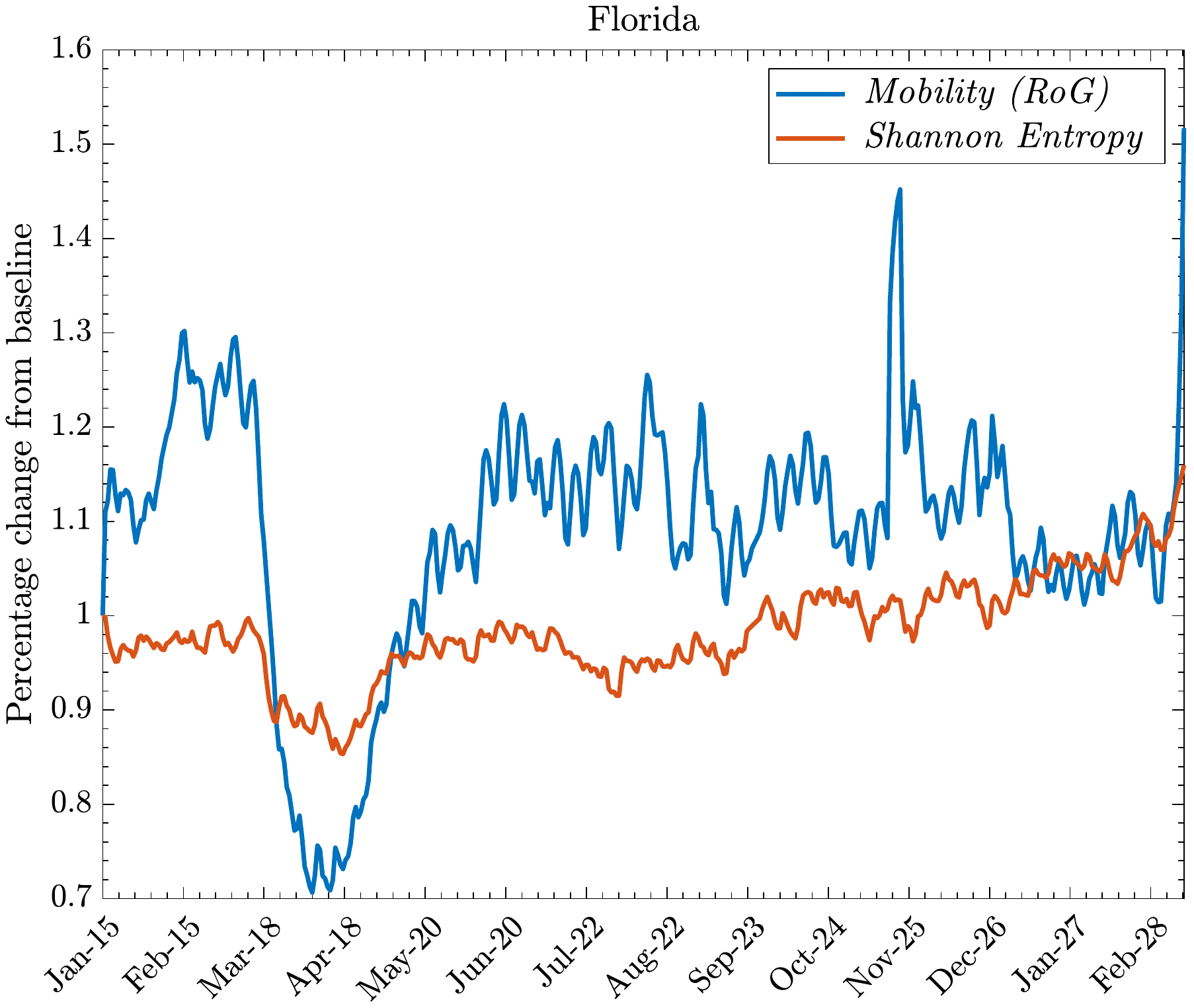}
	\caption{Percentage change of Radius of gyration of individuals' movements and the Shannon uncorrrelated entropy repsect the baseline of Jan 2020. Data courtesy of Camber Systems \citet{camber}.}	
	\label{fig_rog}  
\end{figure}

As regarding with mobility variable, the median radius of gyration in meters of devices which stayed in one location overnight. This metric provides a summary of travel that incor­porates both the number of trips and the distance of every trip. The radius of gyration for user $u$, is calculated first taking the root mean squared distance of a user's movement across space over a given time window from their center of gravity:
\begin{equation}
r_g^{(u)}=\sqrt{\frac{1}{L_i^{(u)}}\sum_{i=1}^{L_i^{(u)}}(r_i^{(u)}-\bar{r}^{(u)})^2}
\end{equation}
where $r_i^{(u)}$ represents the $i=1\ldots L_i^{(u)}$ positions recorded for the device $u$ and  $\bar{r}^{(u)}$ is the center of mass of the trajectory.  For each user, $r_g^{(u)}$ is interpreted as the characteristic distance traveled by user a when observed up to time. Let us notice that the radius of gyration  in a pure diffusive erratic movement should follow  $r_g(t) \sim \sqrt{t}$ in the short time period within a day \citet{liao2019individual,liu2018temporal}. The individual radius of gyration is different from the average travel distance, because an individual moving in a comparatively confined space will have a small radius of gyration even a large distance is covered. On the other side, the radius of gyration can be larger than distance traveled if someone travels with small steps but in a fixed direction or in a large circle.
To calculate the aggregated radius of gyration $RoG$ for a group of devices in a geohash, for every user $u$, one generates their home region $A$ as the region in which they spend the most time in their location set. Then, aggregate this value across a population in a given region and provide an average and percentiles. This is the metric used in CamberSystem \citet{camber} database.
In practice, the radius of gyration represents the vital diameter within which the user is most likely to be found in the observation period and it represents a way to describe human mobility as an aggregated measure of social movements. Since a low radius of gyration corresponds to locations that are close to each other, meanwhile   a large radius of
gyration corresponds to locations which are far apart from each other.  Moreover, people who live in cities may have a lower radius of gyration because they are covering less ground. People who live in rural areas may have a higher radius of gyration because they travel greater distances to achieve the same goals.
In Fig.\eqref{fig_rog} we plot  the percentage change of both  the radius of gyration and the shannon entropy for some of the US states. It is clear how despite the fact that they looks similar, they have pretty different trends showing how they are correlated except for certain omitted variables, which we consider to consist in population density and inter-personal distance. We have already mentioned that these last two variables are difficult to evaluate on a daily basis and only rough estimation are possible.

\begin{figure}[!ht]
	\centering
	\includegraphics[angle=0,origin=c,width=0.6\linewidth]{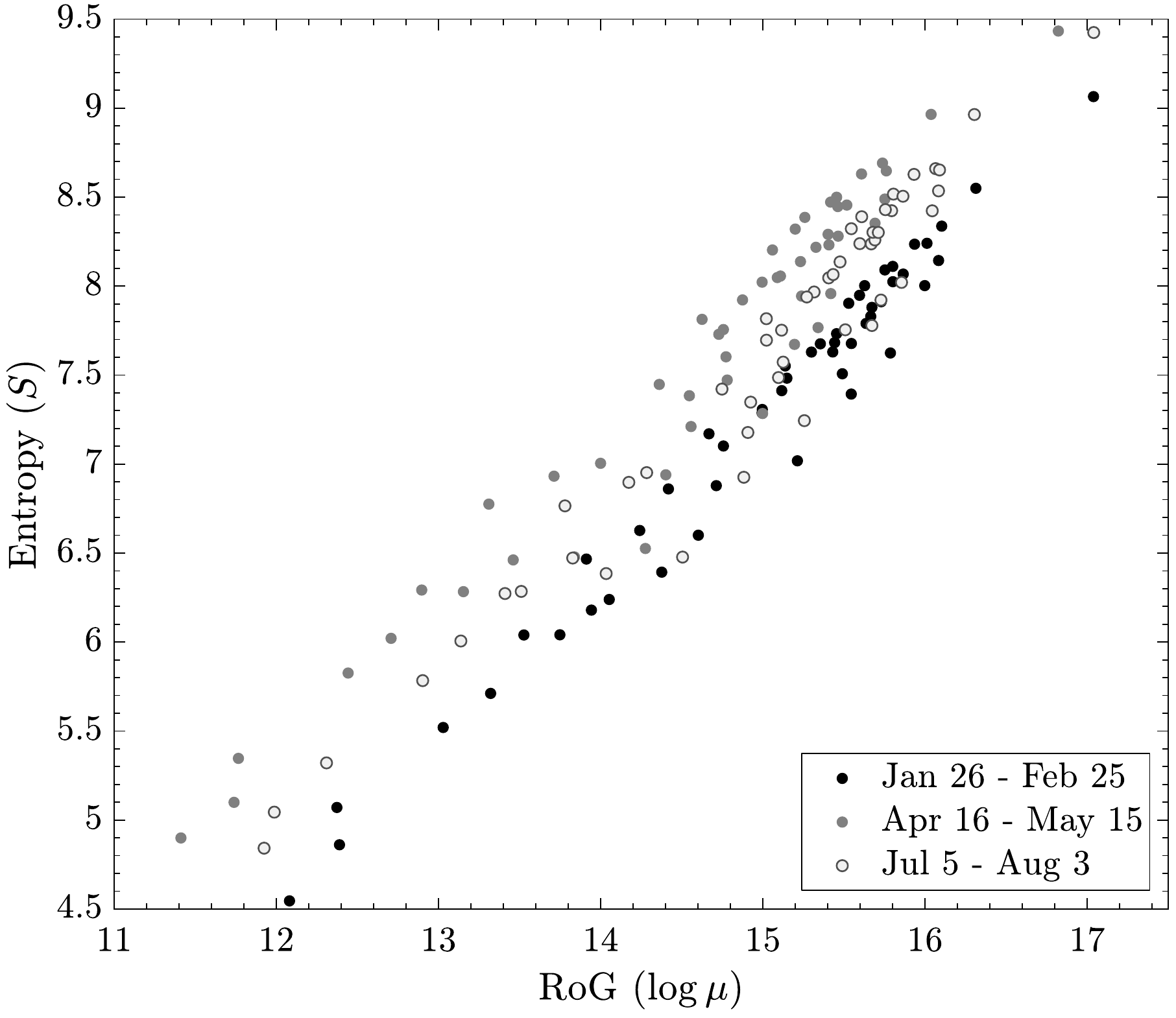}
	\caption{ Panel regression analysis with three dummy variables related to temporal fixed effects in different periods of the year 2020.}	
	\label{fig_panelscatter}  
\end{figure}

Mobility data is the only reliable predictor for the entropy variable, since the population density $\delta $ and individual space $\delta_1$ are sloppy variables which cannot be measured with the same precision and frequency. So we can write the Shannon entropy in terms of mobility  to be related by the following linear regression relation with time specific fixed effects, which omitted variable bias  caused by excluding unobserved variables (population density and proxemic space) that evolve over time but are constant across entities:
\begin{equation}\label{eq_muS0}
S_{i,t} =\beta_0 + \beta_1 \log \mu_{i,t} +\sum_{j=2}^{T}\tau_jB_{j,t} +\varepsilon_{i,t}
\end{equation}
where  $\tau_j$ is the coefficent on time specific dummy variable $B_{j,t}$, the latter equal to one at year $j$, zero elsewhere. In our example $T=3$, for three different periods of 30 days, as reported in Fig.\ref{fig_panelscatter}, namely before the pandemic, during the first outbreak and, last, during the summer of 2020. This model eliminates omitted variable bias caused by excluding unobserved variables that evolve over time but are constant across entities. The constants are in practice evaluated from data since there is no straightforward interpretation of all the variable at play.  So we can perform panel regression analysis for three periods of time over all the $50$ US states as in Tab.\ref{tab_regressmuS0}, that reveals the effects of time specific dummy variables and the strong (cor)relation between Entropy and human mobility as predicted by the theoretical prediction eq.\eqref{eq_entropymobi}. 
\begin{table}[!ht]
	\centering
	\begin{tabular}{l|ccc}
		& \textit{Estimate} & \textit{S.E.} & \textit{tStat}  \\ 
		\toprule 
		$\beta_0$  & -5.93$^{***}$ & 0.23 & -25.37  \\ 
		$\beta_1$  & 0.88$^{***}$ & 0.01 & 56.91  \\ 
		$\tau_2$ & 0.64$^{***}$ & 0.04 & 15.01  \\ 
		$\tau_3$ & 0.32$^{***}$ & 0.04 & 7.62  \\ 
		\bottomrule 
	\end{tabular}
	\caption{Panel regression analysis for 3 periods of time such as before the outbreak, during the first wave, and the end of the first wave. 
		Number of observations: 150, Error degrees of freedom: 146
		Root Mean Squared Error: 0.21
		R-squared: 0.96,  Adjusted R-Squared 0.96. $***$ indicates a p-value<0.001.}
	\label{tab_regressmuS0}
\end{table}
In practice one can evaluate the population density variables $\log \delta_1/\delta$ as from the intercepts from regression analysis as in Fig.\ref{fig_panelscatter}. In theory, it is possible to use the fixed effects regression analysis to  remove omitted variable (i.e. population density) bias by measuring changes within groups across time, usually by including dummy variables for the missing or unknown characteristics. We can see how the density variable has increased the regression intercept after the Covid-19 outbreak, but then going closer to the original value during the summer where gathering restrictions released, since the perceived density increased.

\subsection{Statistical analysis of economic trends}
The economic consequences as the results of restrictions have included increased unemployment, firm closure, stock market volatility and strain on government public finances. Typically, the epidemic as a negative shock on the labor force, so the number of Covid-19 incidence has a negative feedback on the dynamics of economic activity.
We are interested in analyzing human mobility behaviors and how they might be potentially related to short-term regional income, employment rate and other socio-economic factors, \cite{bonaccorsi2020economic,weill2020social,pepe2020covid}. We investigate the correlations between the aggregated mobility and entropic measures against the five  socio-economic indicators.
As already discussed by \citet{pappalardo2016analytical}, we show how the Shannon entropy reveals stronger correlations with some socio-economic indicators than genuine mobility (in terms of  radius of gyration), this can be due by the fact the entropy is a variable which embodies more aspects of social movement trends including mobility.
Entropy, in fact, can be seen as more related to decision making choices than the simple mobility indicator, since it represents an aggregated measure of regularity  respect the ongoing socio-economical patterns. 

In our preliminary study we select some daily and monthly regional economical indicators which have been already analyzed in the research of the short-term  impact of Covid-19 epidemic on the economy \citet{flaschel2021pandemics,reissl2021assessing,guerrieri2020macroeconomic,congress} namely:  employment, consumer spending, electricity production, firms revenues and the coincidence index. We assume that during the Covid-19 crisis the mobility restriction and all the other social distancing interventions  are considered as a shock to the economic activities, so that tahy could be considered the main drivers of  economic trends. At this point, we can assess the statistical measure which determines the association of those indicators against mobility and entropy variables for each state. We do that measuring the linear relationship through the correlation coefficient so we can highlight the aggregated extent to which  each of the economic time series  move together with the mobility and entropy. We show that  some of the economic series are stronger correled with the entropy respect with mobility. Essentially, we think, it can be due to the different content of information in each economic variable and to the fact that entropy embodies a more complex information than just the mobility that for some economic indicator can be an advantage for other indicators it is not. 
\begin{figure}[!ht]
	\centering
	\includegraphics[angle=0,origin=c,width=0.8\linewidth]{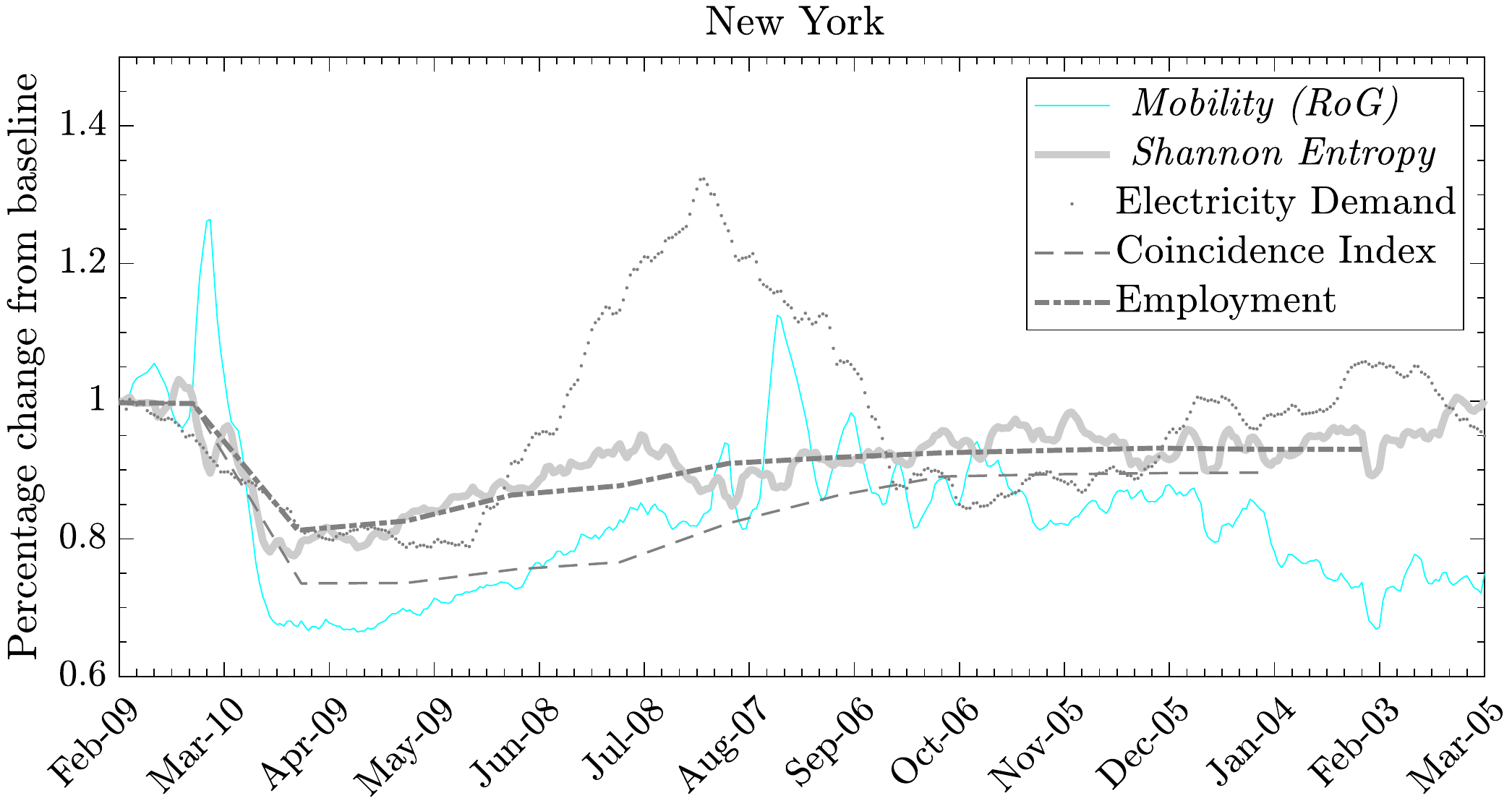}
	
	\vspace{0.5cm}
	\includegraphics[angle=0,origin=c,width=0.8\linewidth]{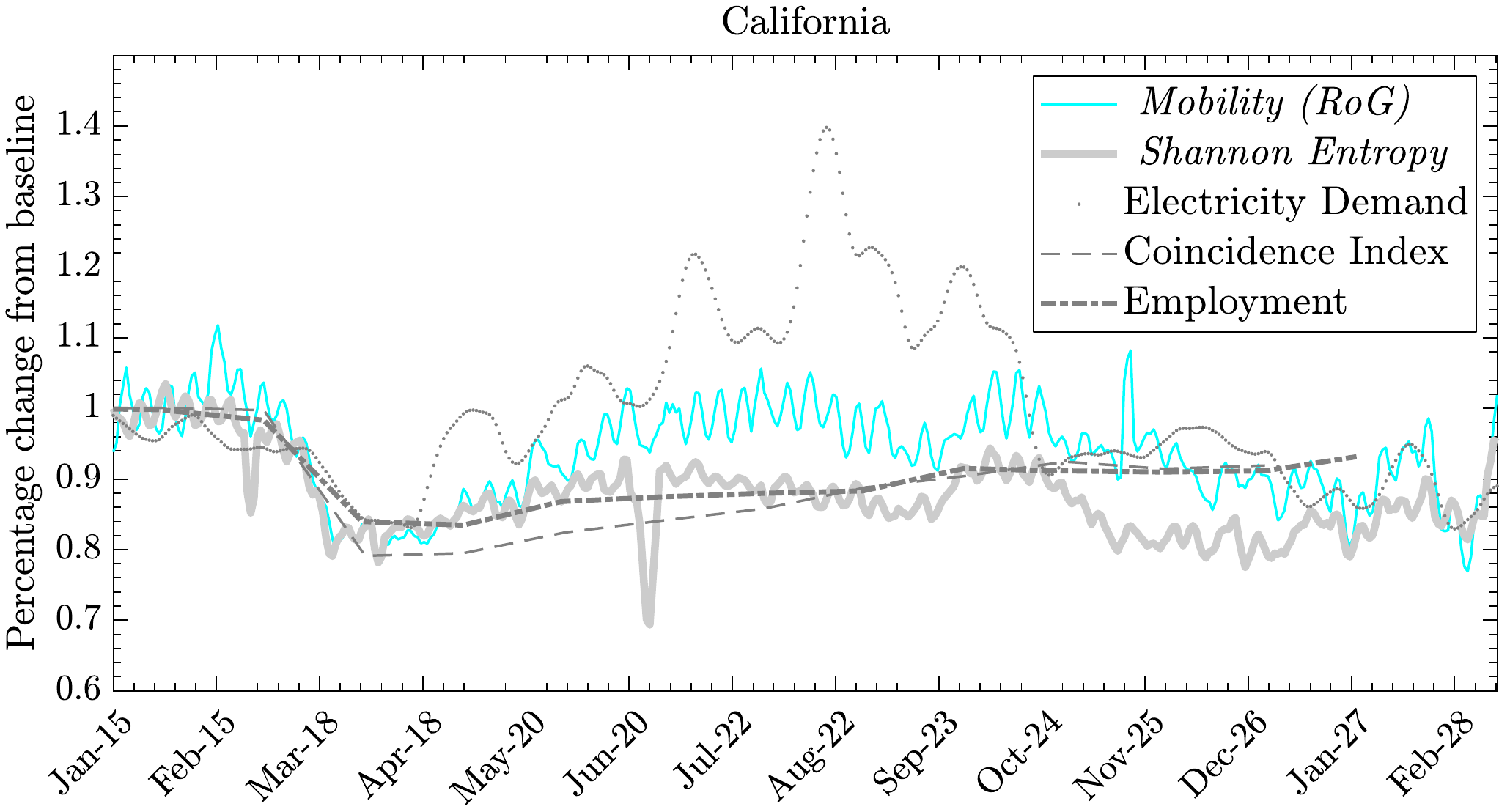}
	\caption{The economic indicators are monthly reported. Data courtesy of Camber Systems \citet{camber}, U.S. Bureau of labor statistics  \citet{ibs}, Federal Reserve Bank of Philadelphia  \cite{federalphi} and Electricity demand EIA \cite{iea}, for mobility, labor, economic and energy data respectively.}	
	\label{fig_economic_entropy}  
\end{figure}
As regards with the impacts on labor market, employment  rates  in  the  United  States  fell  dramatically  during the first months 2020  as the  repercussions  of  the  COVID-19  pandemic  reverberated  through  the  labor  market.  
However, the pandemic-related economic pause and lockdown differentially affected the employment opportunities of persons working in different sectors. Workers whose jobs could be performed remotely from home, continued to work from their home office, meanwhile,  workers who provided essential services (health care professionals and grocery store clerks) kept their usual work routine. In particular, regions with economies that rely on the movement of people (like  tourism) faced substantially higher unemployment at the end of 2020 than regions with core industries based on the movement of information.
Population mobility is strictly related to consumer decisions about what to buy, how much to buy, and when to buy among many goods and services. They do not only satisfy their own needs, but  also  determine  how  much  of  which  goods and services ultimately will be produced. The production of these goods and services creates jobs in all sectors of the economy. 
As regard with  consumer spending perspective, we have picked the only high frequency data which has to do with electricity demand.  Business closures and changes to normal routines related due to policy restrictions have caused electricity demand to decrease in March and April compared with expected demand, after accounting for seasonal temperature changes. Such indicator measures  how much electricity each end-use sector consumes and the varying effects of COVID-19 mitigation efforts on the sectors. We have used data of demand for electricity as from reoprted by IEA \cite{iea} of regional electricity production in megawatt-hours units. 
Ultimately, we have selected monthly coincident index for each of the 50 state as produced by the Federal Reserve Bank of Philadelphia \cite{federalphi}.
The coincident indexes combine four state-level indicators to summarize current economic conditions in a single statistic. 
These indexes are monthly indicators of economic activity for each of the 50 U.S. states, based on a composite of four widely available data series on state conditions: total nonfarm payroll employment, the unemployment rate, average hours worked in manufacturing, and real wages and salary disbursements.

In Fig.\ref{fig_economic_entropy} we show how economic indicators seems to be better aligned with entropy rather then the mobility especially for US states as New York state and other US states where entropy and mobility have opposite trends at the beginning of 2021. In states, as California, where the two mobility indicators have similar trends, the correlations between the mobility and the economic indexes are more indistinguishable. However, due to the complexity of the subject, our empirical exercise represents only a qualitative and preliminary investigation since we have focuses our research study on setting a robust and quantitative description of population mobility, so putting the ground for further and more rigorous studies which can connect socio-economical indicators to mobility ones.

\begin{figure}[!ht]
	\centering
	\begin{subfigure}[c]{0.3\textwidth}
		\centering
		\includegraphics[ trim = 0 0 0 5cm,clip, angle=0,origin=c,width=0.99\linewidth]{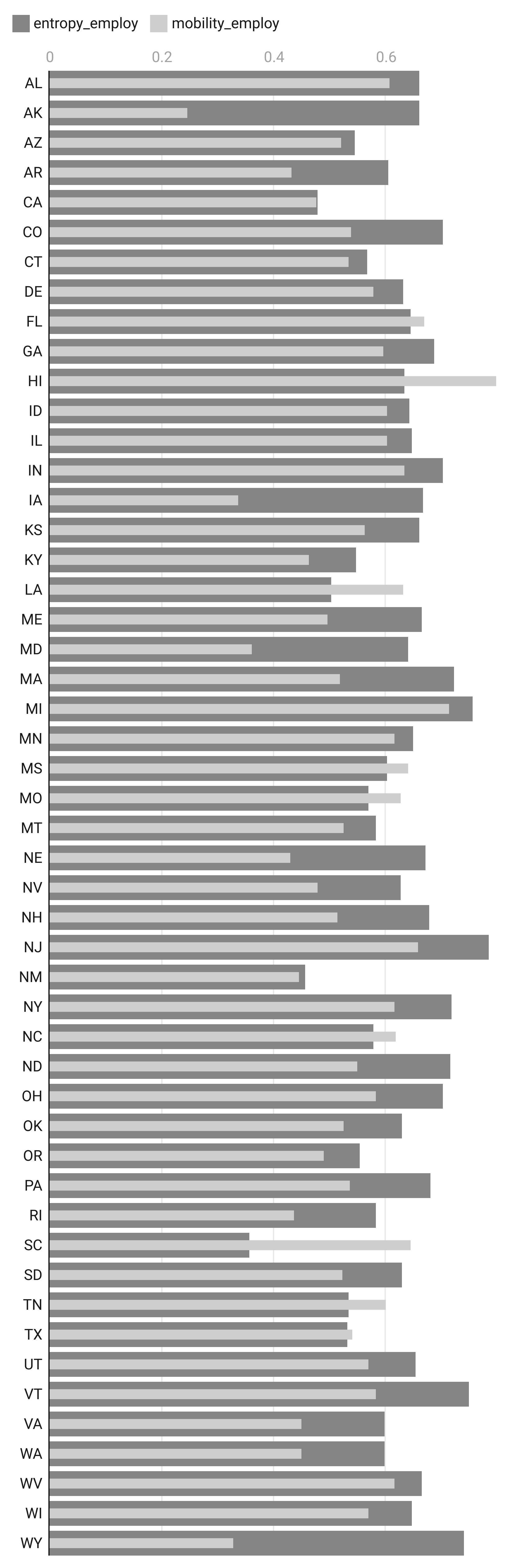}
		\caption{Employment }
	\end{subfigure}
	\begin{subfigure}[c]{0.3\textwidth}
		\centering
		\includegraphics[ trim = 0 0 0 5cm,clip, angle=0,origin=c,width=0.99\linewidth]{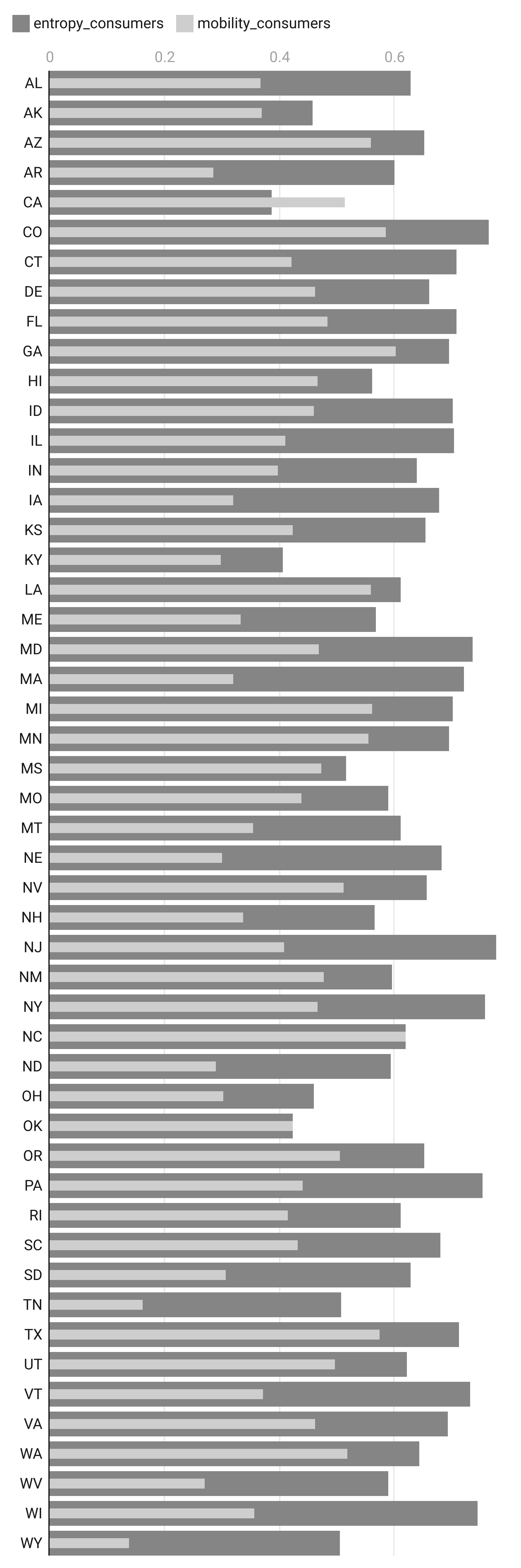}
		\caption{Consumer Spending }
	\end{subfigure}
	\begin{subfigure}[c]{0.3\textwidth}
		\centering
		\includegraphics[trim = 0 0 0 5cm,clip,angle=0,origin=c,width=0.99\linewidth]{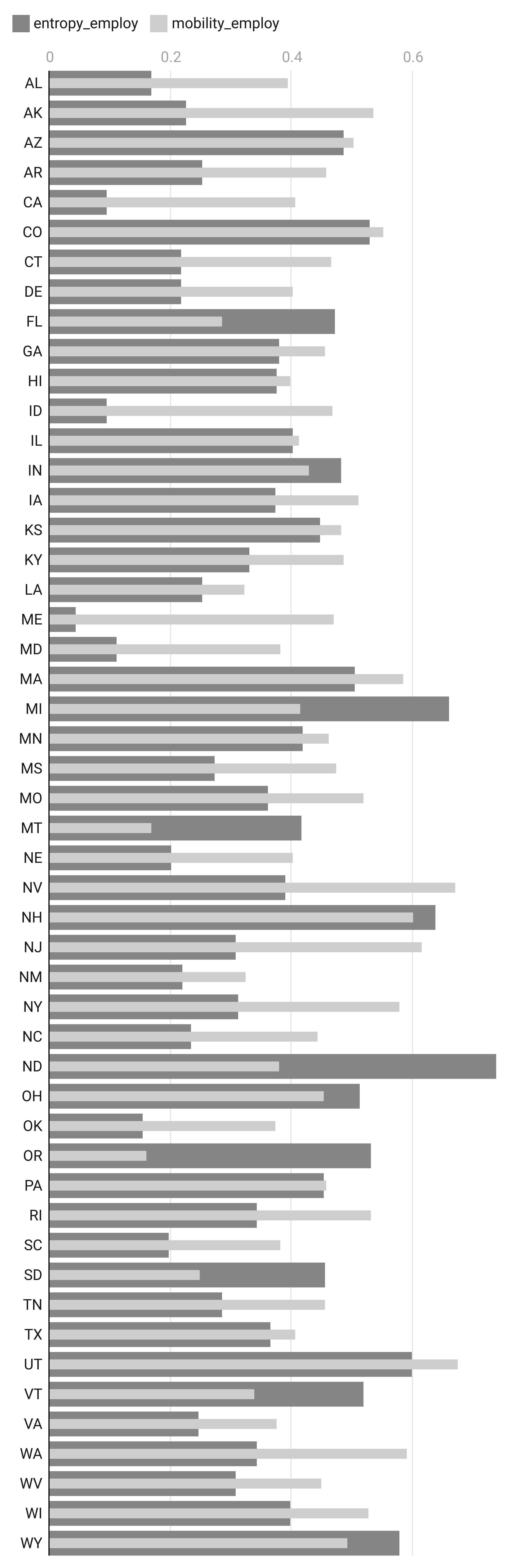}
		\caption{Energy production  }
	\end{subfigure}
	\caption{Correlation coefficient for each state between the economic indcators versus the mobility (light thin bars) and uncorrelated entropy (dark thick bars) for two different economic indicators: (a) employment rate, (b) consumer spending and (c) energy consumption.  Longer bars indicate a stronger correlation between the time series.}	
	\label{fig_correconomic_entropy}  
\end{figure}

%
%
At this point we want to measure the relationship between each of the selected economic indexes versus the mobility and entropy variables for each of the 50 US states as reported in Fig.\ref{fig_correconomic_entropy}. We see how the entropy is systematically highly correlated to employment and coincidence index than the mobility in almost all the states. Differently, mobility (as RoG) is more correlated with energy demand. This is summarized in the average correlations as reported in table Tab.\ref{tab_correlations50} where we have computed the median  correlations and their confidence intervals. We observe that Entropy shows a significantly higher degree of correlation against employment and coincidence index than mobility. This can be due to a more complex nature of the entropy which accounts for mobility together with other social distances measures as indicator of regularity of locations patterns.  Meanwhile mobility is more strongly correlated with energy demand, and this can be due to the fact that the consumption of electricity is more sensitive to a change in movements of individuals rather then some sort of regularity of those.

\begin{figure}[!ht]
	\centering
	\begin{subfigure}[c]{0.3\textwidth}
		\centering
		\includegraphics[ trim = 0 0 0 5cm,clip, angle=0,origin=c,width=0.99\linewidth]{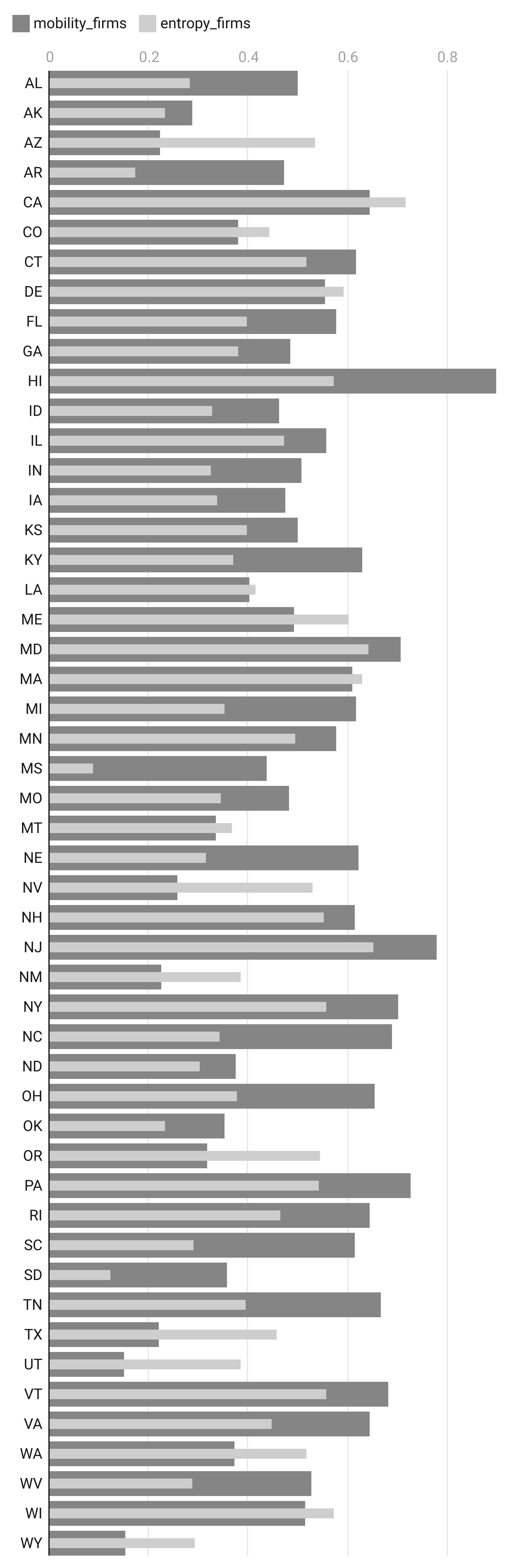}
		\caption{Firm Revenues }
	\end{subfigure}
	\qquad \qquad
	\begin{subfigure}[c]{0.3\textwidth}
		\centering
		\includegraphics[trim = 0 0 0 5cm,clip,angle=0,origin=c,width=0.99\linewidth]{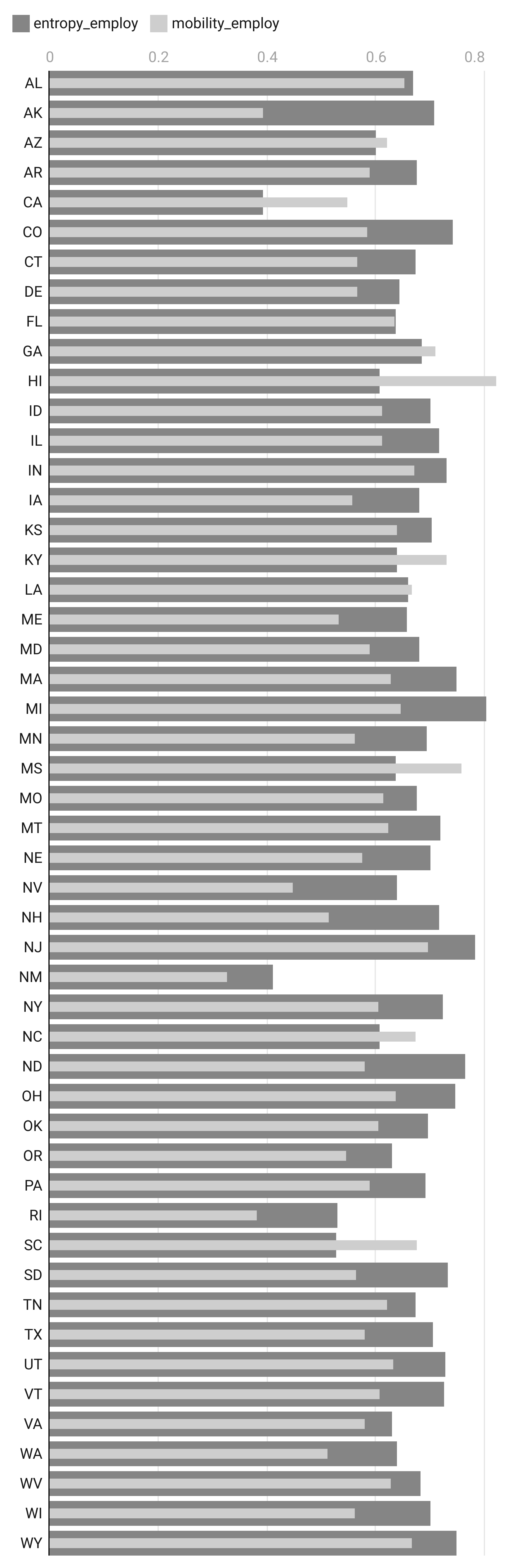}
		\caption{Coincidence Index  }
	\end{subfigure}
	\caption{Correlation coefficient for each state between the economic indcators versus the mobility (light thin bars) and uncorrelated entropy (dark thick bars) for two different economic indicators: (a) Firms revenues and (b) Coincidence index. Longer bars indicate a stronger correlation between the time series.}	
	\label{fig_correconomic_entropy2}  
\end{figure}

\begin{table}[!ht]
	\centering
	\begin{tabular}{l|cc| cc}
		\textit{Indicator} & Mobility  & \textit{C.I.} &   Entropy  & \textit{C.I.}   \\ 
		\toprule 
		\hline
		Employment  & ${\textit{0.55}}$ & {\normalsize [0.43, 0.64]} &  $\textbf{\textit{0.64}}$ &{\normalsize  [0.53, 0.72]}  \\ 
		\midrule
		Consumer Spending  & ${\textit{0.43}}$ & {\normalsize [0.29,0.56]} & $\textbf{\textit{0.65}}$  & {\normalsize [0.48, 0.74]}  \\
		\midrule
		Energy Production  & $\textbf{\textit{0.46}}$ & {\normalsize [0.32,0.59]} & $\textit{0.36}$  & {\normalsize [0.16, 0.56]}  \\ 
		\midrule
		Firms Revenue  & $\textbf{\textit{0.51}}$ & {\normalsize [0.24,0.69]} & $\textit{0.34}$  & {\normalsize [0.26, 0.60]}  \\ 
		\midrule
		Coincident index & $\textit{0.61}$ & {\normalsize [0.51, 0.69]} & $\textbf{\textit{0.69}}$& {\normalsize [0.61, 0.75]}  \\ 
		\bottomrule 
	\end{tabular}
	\caption{Correlation coefficient among all the 50 US states. Median values of the correlation coefficient and their 10th and 90th percentiles.}
	\label{tab_correlations50}
\end{table}



\clearpage
\subsection{Validity and limitations of model assumptions}
\begin{figure}[!ht]
	\centering
	\begin{subfigure}[c]{0.47\textwidth}
		\centering
		\includegraphics[angle=0,origin=c,width=0.75\linewidth]{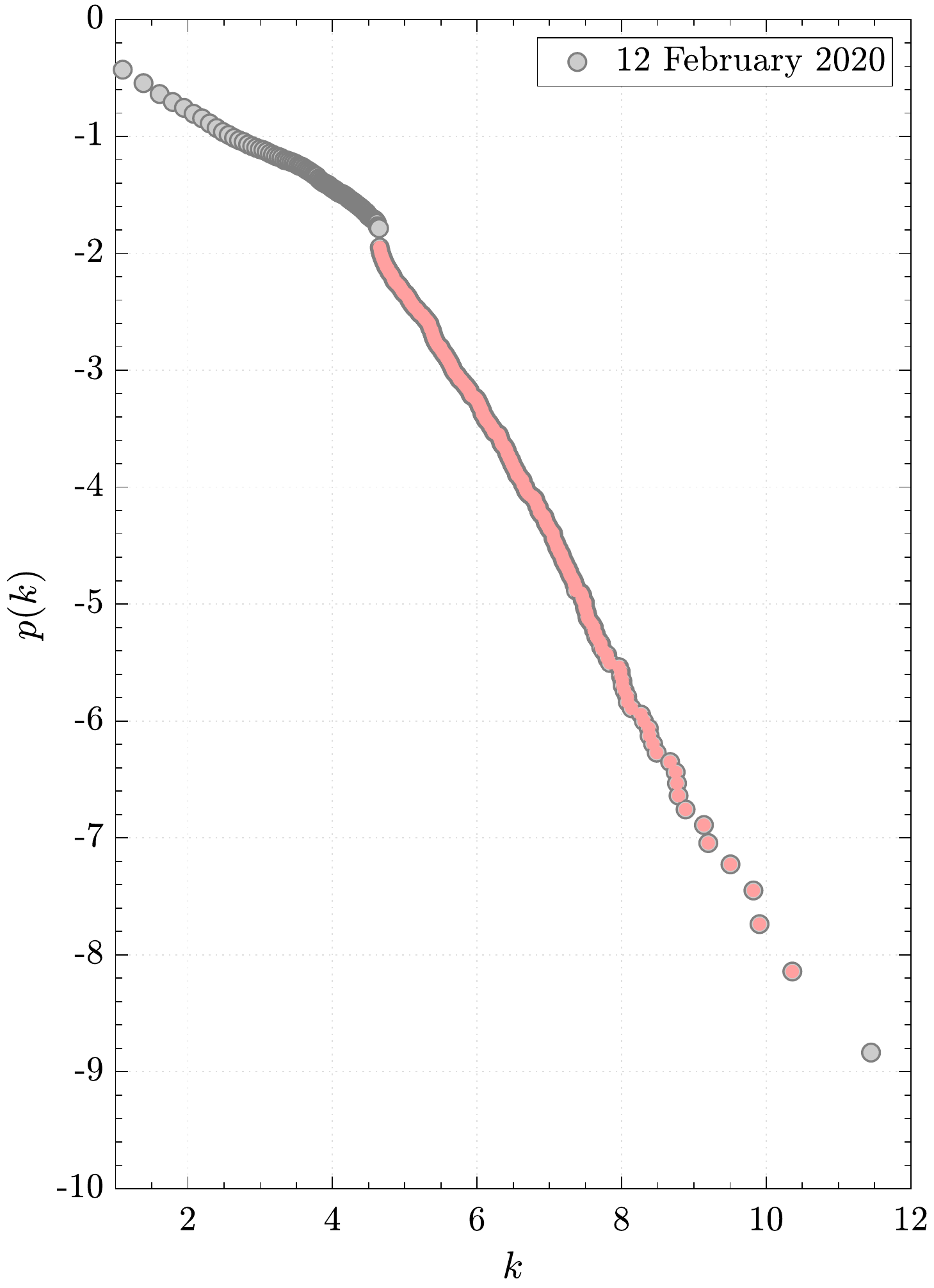}
		\caption{}
	\end{subfigure}%
	\begin{subfigure}[c]{0.47\textwidth}
		\centering
		\includegraphics[angle=0,origin=c,width=0.77\linewidth]{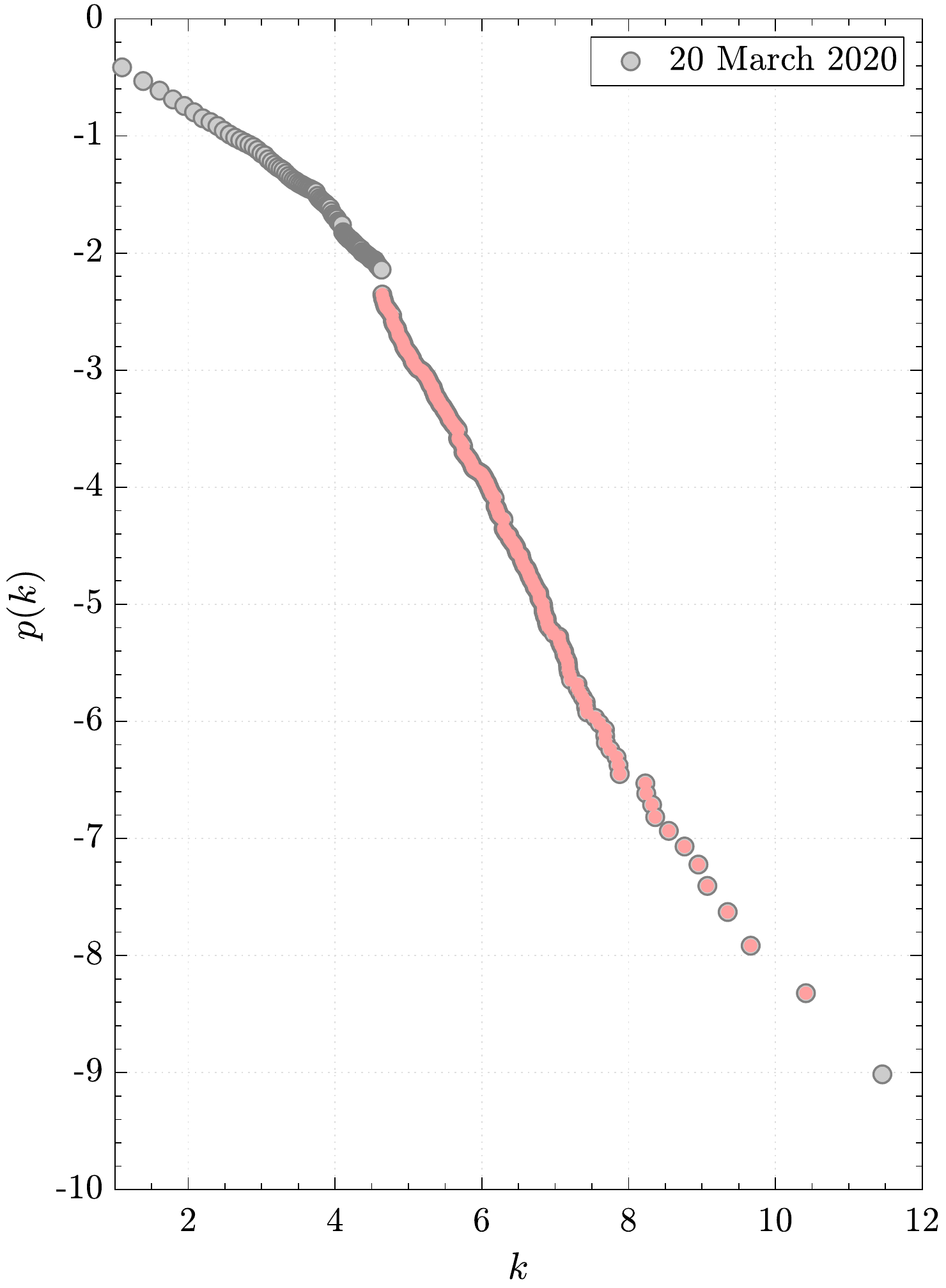}
		\caption{}
	\end{subfigure}%
	\\ \vspace{0.5cm}
	\begin{subfigure}[c]{0.47\textwidth}
		\centering
		\includegraphics[angle=0,origin=c,width=0.75\linewidth]{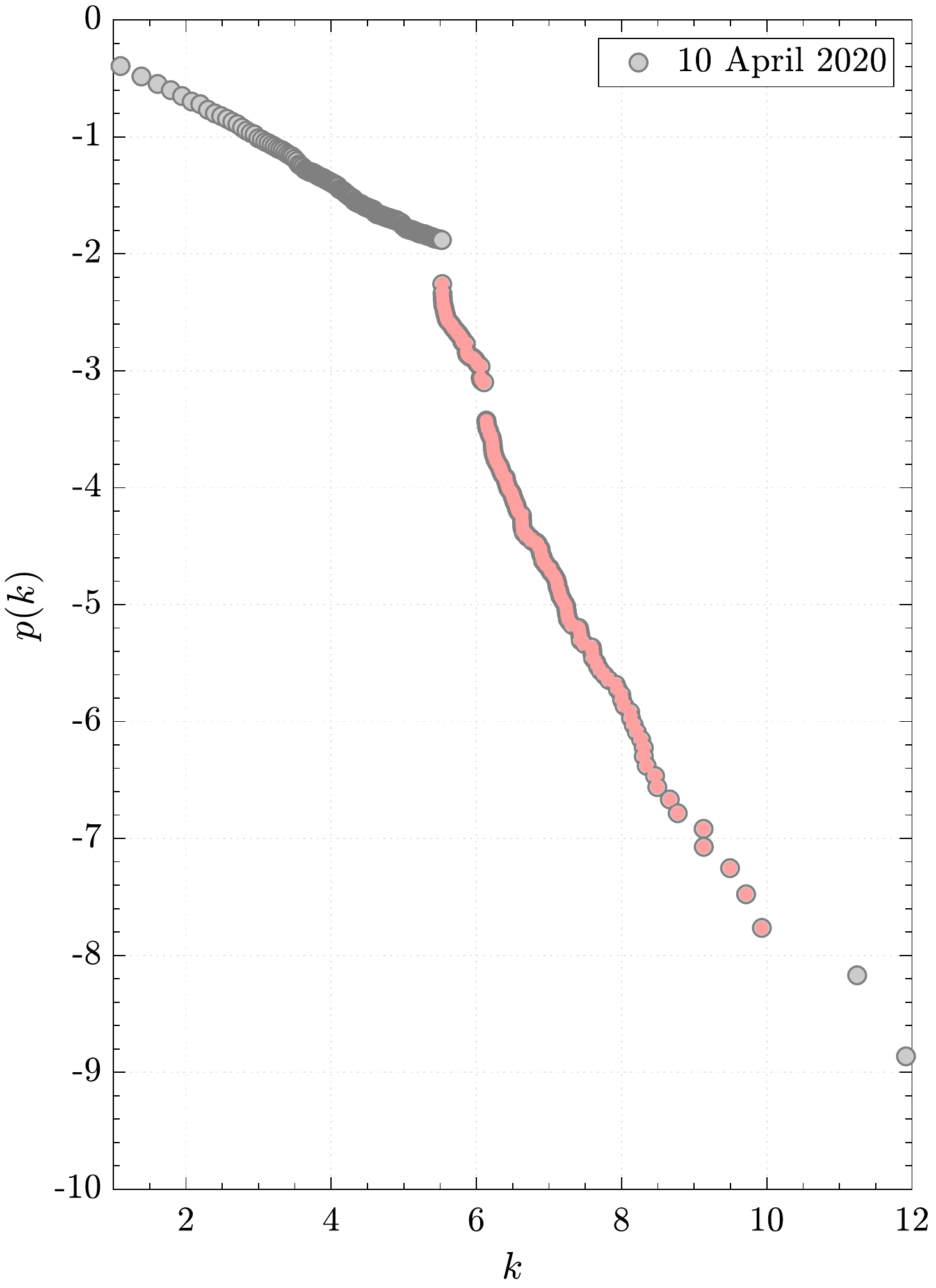}
		\caption{}
	\end{subfigure}%
	\begin{subfigure}[c]{0.47\textwidth}
		\centering
		\includegraphics[angle=0,origin=c,width=0.75\linewidth]{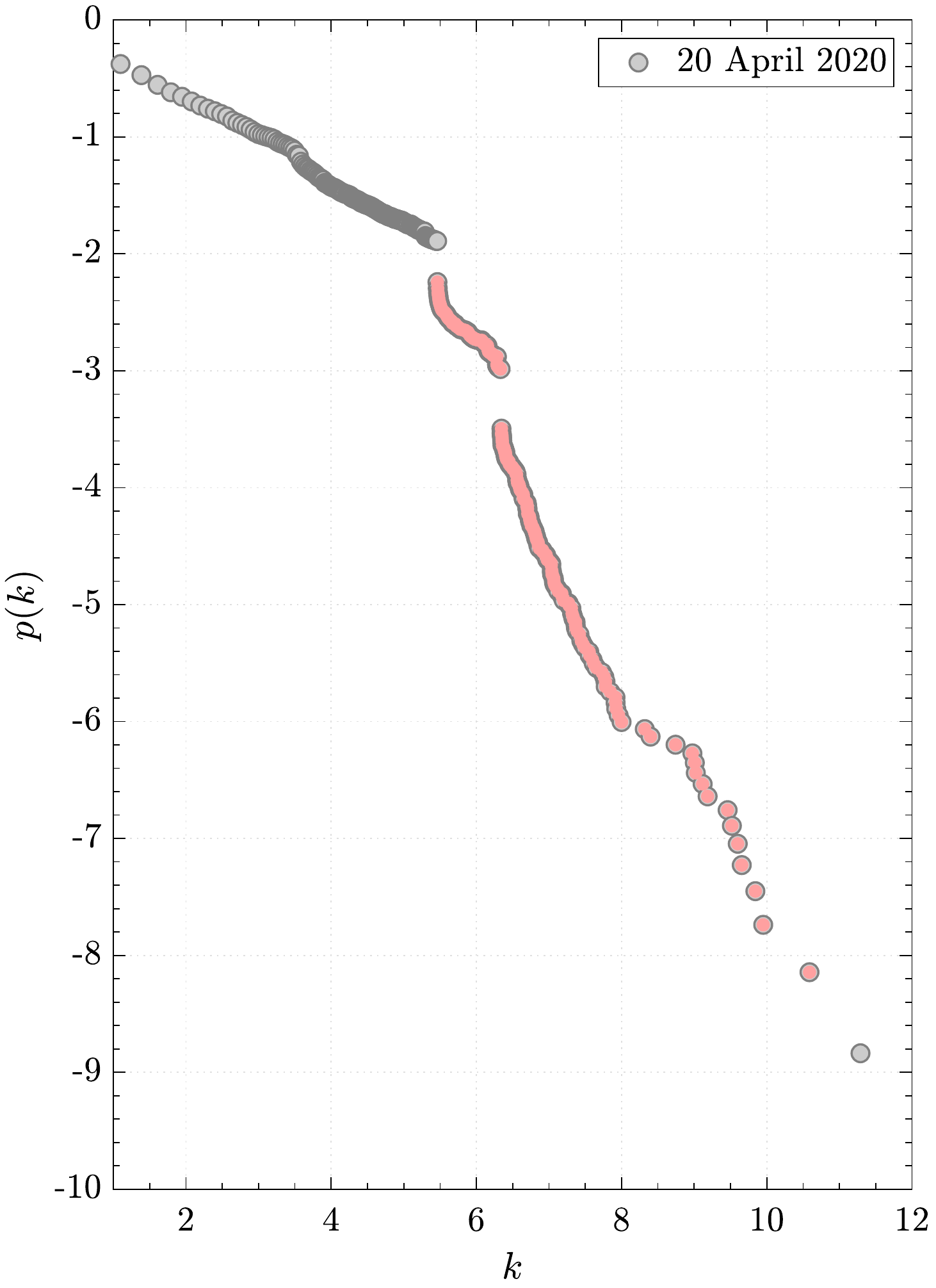}
		\caption{}
	\end{subfigure}%
	\caption{Complementary cumulative function of the frequency of visiting  in a log-log plot. Using real world data by Safegraph \citet{safegraph} for New York city county for the dates on the legend. So that visiting probability $p(k)$ has a clear fat tail distribution which can be represented through a generalized Pareto distribution.}	
	\label{fig_freq_visiting}  
\end{figure}

Let us now go into a more fine-grained analysis of the visiting pattern structures, namely the probability distribution of the number visits that each census block have received in a day. In order to obtain a more realistic estimate for the visiting probability $p(k)$, we have needed to use SafeGraph \citet{safegraph} Data consortium, which provides the number of devices in census block group that stopped in the given destination census block group for $>1$ minute during the time period. In Figs.\ref{fig_freq_visiting} we show how the distribution has changed during the epidemic period, groing from a clear power law tailed distribution in pre-pandemic period to a more disrupted tail where reorganization of interaction patterns has affected  the distribution of how locations are visited (as also distances traveled).  

From the previous empirical results on mobility, we could assume a generalized Pareto distribution for the visiting locations:
\begin{equation}
p_k=\lambda\left(  1+\dfrac{k-k_{0}}{\alpha\lambda}\right)^{-(1+\alpha)} \qquad \text{ for }\; k>k_0\geq 1, \alpha>0 ,\lambda>0 
\end{equation}
which shows power law tails with slope of $\alpha$, such distribution has finite mean when $\alpha >1$ and it has finite variance when $\alpha>2$.

Let us notice that for the particular case when $1/\alpha=0$ the probability density becomes exponential $p_k=\lambda e^{-\lambda (k-k_0)}$.
In the generalized case, the Shannon Entropy is:
\begin{equation}
S_\mathbb{U}^{(gp)}=1-\log \lambda +1/\alpha  =1/\alpha + S_\mathbb{U}^{(exp)}
\end{equation}
so larger than the exponential distribution entropy.

In our theoretical representation we miss two important real life evidences, first a spatial heterogeneity is much larger than the one represent in our model but captured by the empirical Shannon entropy from data, as discussed in the appendix. Second we miss both a theretical and emprical temporal heterogeneity of individuals' paths.
The current mobility models are based on an assumption that human movements are randomly distributed in
space and time, hence are approximated by a Poisson process,  however recent studies have swhon the non-Markovian nature of population movements, thus possessing a memoryless structure.

Respect to the random diffusion of particles, human daily travel has a higher degree of regularity, so an higher predictability of individual movements \citet{song2010limits}. This happens since individuals in urban travel are purposeful and socially contextualized, so they have different destination choice strategies.
In particular, studying the erraticity of people movements gives a better understanding of the limitations of our model and consequently in a better interpretation on the use of data like mobility and  social distancing, as well as the necessity to calibrate the model calibrating such discrepancies. A crucial point while using mobility data is that we assume they are a good proxy for relative velocity of people in the erratic movements in the collisional framework. In order to be true, we essentially assume that the movements of individuals are uncorrelated so that the mean relative velocity among colliding individuals is proportional to their absolute speeds (Maxwell-Boltzmann random mean square velocity  condition). From the kinetic theory of gases, the general form of the probability density function of the speed is of the form $p(v)=cve^{-cv^2/2}$ where $c$ is the inverse of the variance that is related to the concept of temperature derived from the kinetic energy in the collisional theory (namely the 2D Maxwell-Boltzmann Distribution). Let us observe that using SafeGraph database we make use of another proxy for mobility which is the evaluation of distance traveled  metric, so measuring the amount of movement occuring within a population, as reported in SafeGraph database \citet{safegraph}. Despite  RoG and distance-traveled measures are not totaly equivalent, they are considered to be both proxies of human mobility since  probability of displacement  and frequency of the radius of gyration show  almost identical trends \citet{hawelka2014geo}. In our theretical framework, distance-traveled correspond to the velocity of each individual meanwhile RoG is represented as the mean square velocity. Those quantity are proportional within our collisional model.

\begin{figure}[!ht]
	\centering
	\includegraphics[angle=0,origin=c,width=0.8\linewidth]{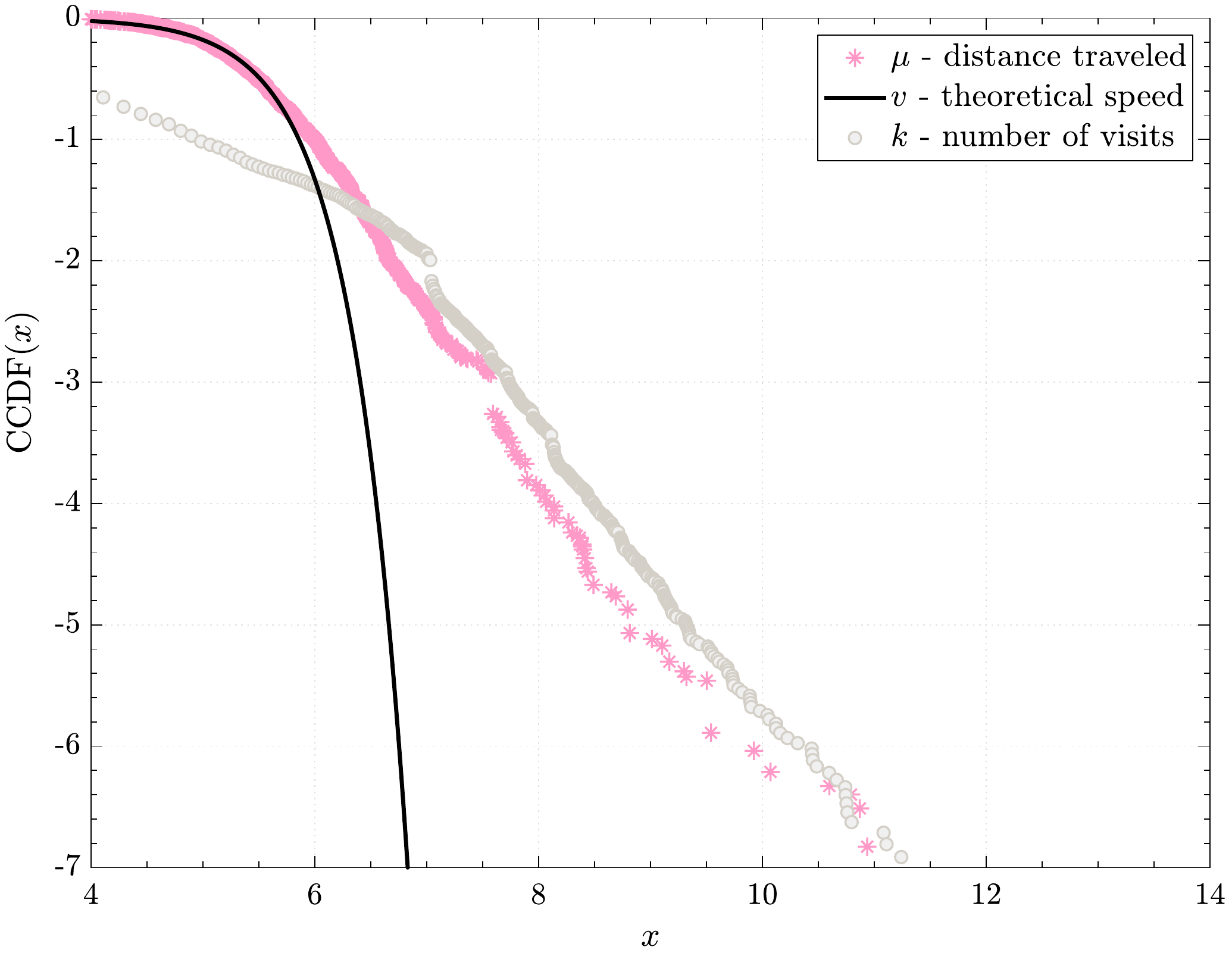}
	\caption{Complementary Cumulative denisty function of different variables $x$ taken from SafeGraph datbase in the case of New York county. Such result confirms the fairly good approximation of the ergodic assumption between the distribution of visited locations and speed distribution. By the way the theoretical prediction fails to describe the heavy tail behavior since we only rely on a population with homogeneous behaviors, so  exponential distribution of the speeds. }	
	\label{fig_mobdistribution}  
\end{figure}

At this point we evaluate  the agreement of theoretical speed distribution $p(v)$ against the true distribution of distance traveled per day $p(\mu)$ evaluated from real data as reported in SafeGraph \cite{safegraph}. As reported in Fig\ref{fig_mobdistribution}, the two distributions have similar expected value $E[\mu]\simeq E[v]$ but the show a clear difference in the tail, since the observed complementary cumulative density function of mobility show a clear power law tail in contrast with the exponential trend of the theoretical Maxwell-Boltzmann distribution of velocity. This fact stresses how our aggregate description of a mean field variable ov mobility is well replicated by the collisional model. On the contrary such model does not capture the important mobility heterogeneity among individuals  (as long range travelers).
Another important result is about what we call the ergodic assumption of movements, that essentially comprehends the fact that the movements of individuals are equivalent to the frequency of location visited by those, so that $p(\mu)\sim p(k)$. Such property is asymptotically confirmed in our analysis, for example again in Fig.\ref{fig_mobdistribution} where we see that those probability have identical power law distribution shape, except at short scale where data collection issues are involved. In fact, in SafeGraph repository the number of visits received by each census block is computed indirectly from the information of how many other census blocks have been visited by devices on a considered census block. So we have a poor estimation of places which are visited only few times, because they are included in larger census blocks.

Moreover there are other two intrinsic assumptions in the collisional model: that each individual performs markovian trajectories (memoryless assumption)  and that, a part from direct collisions, each individual does not influence the trajectory of others (short-range interactions). Those assumptions exclude the presence of long range auto-correlations and cross-correlations of each individual trajectory. Considering such effects will require a study of individual trajectories over time and so introducing other definition of information entropy, as the Kolmogorov entropy as discussed in the appendix.
 When the short correlation assumptions are not fulfilled which contribute differently over time to the  mean relative velocity. At this point, observing at Entropy measures we can realize that there is such change in erratic behavior, so the necessity to calibrate again the model in order to make mobility data consistent with our hypothesis in the model.
In particular, in reality if one looks at the distribution of the inter-event times associated with the individual locations  depicting long-tailed temporal distribution, so that  most locations are visited at high periodicity, while few locations encounter long waiting times. At this stage, se do not have data on individual trajectories during the time period of the pandemic.
However, such limitations are mainly important at the level of individual paths. In our mean field study, we have focused on aggregate mobility so that are analytical predictions are fairly  effective in describing observed overall mobility patterns.

\section{Conclusions}
The efforts to mitigate the Covid-19 pandemic has forced many governments to impose lockdown policies for the population.  However, such mobility restrictions have come at significant social and economic cost.
The analysis of entropy measures of mobility patterns of real-time datasets represents a promising reliable and effective tool for monitoring human mobility patterns. This plays an important role in studies exploring migraion flows, tourist activity  as well as for examining the spread of diseases and for epidemic modeling. Moreover,  home confinement and restrictions on production have affected the behavior of individuals and economic activity leading to a shift in typical human behavior resulting in a crumple of economic predictions. 
Individuals' interpersonal distances, human mobility and population density are the most important factors which can either increase or decrease the contagion diffusion of an airborne transmissible disease as the Covid-19 and at the same time  they have an impact of preferences for products purchase decisions, posing a question of how individual heuristics might form large-scale patterns. In fact,  Covid -19 pandemic,  along with the associated lockdowns, mobility restrictions and physical distancing rules, has altered labor market structure, production chain and spending patterns of consumers affecting so the supply of and demand for many products. 
Our study has been designed in a theoretical framework where mathematical modeling, statistical learning methods and big data analytic can coexist. 
Specifically, we have studied a possible relation between the definition of information entropy of loction visited by indiviuals with their mobility as well as other factors like population density and interpersonal distances. 
We have also investigated the extent to which mobility and entropy correlate with different economic variables.  Unveiling such correlation may therefore prove useful for the design of region-specific lockdown policies, that balance the epidemic spread and economic losses.
In a short time range framework, the interaction rate $\mathbb{B}_t$ in eq.\eqref{eq_intro}, is important in explaining how the trend of the epidemic reproduction number  is directly related to mobility and proximity which are part  of the notion of Shannon uncorrelated entropy.  On the other side, the lockdown intensity $\mathcal{L}_t$, in eq.\eqref{eq_intro2}, is also strongly related with mobility restrictions and confinement policies, so that the Shannon Entropy is a good candidate to capture the short range economic impact on labor market and consumer spending, where a more comprehensive indicators is required beyond just straight mobility.

%

\section{Acknowledgments}
Fabio Vanni acknowledges support from the European Union's Horizon 2020 research and innovation programme under grant agreement No.822781 GROWINPRO - Growth Welfare Innovation Productivity.
\section{Conflict of interest}
The authors declare that there are no conflicts of interest regarding the publication of this paper.


\bibliographystyle{apalike} 
\bibliography{referencesEE} 
\addcontentsline{toc}{chapter}{Bibliography}

\newpage
\appendix
\appendixpage
\addappheadtotoc

\section{Entropic measures }
We represent a situation where a certain number of individuals are represented with two extreme behaviors. Some of them can freely move with erratic movements others stand still (so completely regular) and out from the interaction process in society. We have presented different types definitions of entropy in terms of the data used. We want know to provide some theoretical evaluations which relies on an epidemiological framework of $N$ non-interacting (uncorrelated) individuals moving randomly in a region of size $L$  and area $A=L\times L$ where each individual occupies a small portion of the area $4\pi r^2$ where $r$ is the collision radius which is larger if an agent maintains a small average inter-personal distance with others. We always assume a sparse condition for which $4\pi r^2 \ll A/N $. 

\subsection{Configurational Entropy and Gibbs picture }\label{app_config}
We assume that each individual has the same velocity and radius of others.
The effective area allowed for a single individual is $A=(L-2r)^2$ and let us observe that each individual has $4\pi r^2$ less area than the previous individuals so that the configurational space is:
\begin{equation}
A_N=\frac{1}{N!}\prod_{n=1}^{N}\left( A - (n-1)4\pi r^2 \right)
\end{equation}
where the factorial is the normalization factor due to the permutation of identical individuals. We have also ignored the samll correction the excluded region around the first individual overlaps the excluded region near the borders of the region.
At this point we can write down the entropy on bits of information which is the Gibbs entropy  under constant mobility expansion so to get:
\begin{align}
S_{\mathbb{Q}} &= -\log A_N \\
&= -\log N! + \sum_{n=1}^{N}\log \big(A - (n-1)4\pi r^2\big)
\end{align}
Consequently, under the sparsity assumption and large number of individuals, one can write the the configurational entropy in terms of Gibss (or random) picture as:
\begin{align}
S_{rand} &=S_{\mathbb{Q}}/N \\
&\simeq 1+\log\Big(\frac{A}{N} -4\pi r^2\Big) = 1+\log\Big(\frac{A}{N} -\frac{A_1}{1}\Big)\\
&\simeq  1+\log\Big(\frac{1}{\delta} -\frac{1}{\delta_1}\Big)
\end{align}
where we have used the Sterling approximation for factorials and the sparse condition. We observe that $A/N = 1/\delta$ is the inverse of the population density, which represents the number of individuals which actively participate to the collision process over the effective region which is explored with the same frequencies in all its parts.
All this assumptions push the previous configurational entropy to be associated the so called random entropy $S_{rand}$ which is equivalent to the Gibbs entropy of equiprobable configurations. Let us notice that in principle the interaction radius $r$ can dependt both on the size of the region and the number of individuals or even depend on cultural attitudes. That is why it would be better to find another variable which is more free from such interdependence and

We can also include the effect of the human mobility by considering the definition of thermodynamic entropy   in $2$ dimension due to a change of temperature $T$ and pressure $P$ in a system with constant kinetic energy as $\frac{\partial S}{\partial A} \big|_E = \frac{P}{T}$
where the energy is purely kinetic so that $E\propto NT$, so keeping a constant energy amounts to keeping a constant temperature.
 So substituting the expression for the configurational entropy $S_{\mathbb{Q}}$ and differentiating, we obtain the important relation among mobility, interpersonal distancing and population density $\frac{A}{N} -2\pi r^2\propto T$.
Then we can  take the temperature to be related to the mean square velocity in $2N$ velocity coordinates as $T= \frac{1}{2}k \mu ^2$, so that the random entropy is proportional to mobility.
However the dependence of the random entropy on velocity is dependent on variables that we do not have control about, we are not able to follow the change of entropy over time just starting by the mobility. In order to fix such limitation of the Gibbs picture, we can generalize the calculation of the entropy, assuming independently the existence of a give density and mobility separately.  
As a consequence, in order to detect a more clear relation between entropy and mobility other than the population density, we use a different approach which account for a large phase space for both position and momentum.

\subsection{Shannon Entropy and Boltzmann picture}\label{app_bolt}
Now let us discuss the case where individuals are allowed to have a velocity distribution different from the previous Dirac delta distribution as in the configurational entropy above.
We consider $N$ individuals in a $2$ dimensional region. Canonical coordinates and momenta are $\boldsymbol{p}=(p_1\ldots p_{2N}),\boldsymbol{q}=(q_1\ldots q_{2N})$ so that every point in the $4N$ dimensional phase-space $k(\boldsymbol{p},\boldsymbol{q})$ corresponds to a possible state of the social system. The volume measure of the phase-space is defined as:
\begin{equation}
d\Gamma =\prod_{i=1}^{2N}\frac{dp_i dq_i}{c}
\end{equation}
 where $c=(m^2\mu_1^2\pi r^2)^N$ is a dimensional constant.
 The probability for the system to occupy the state at point $\boldsymbol{k}$ is $\rho (\boldsymbol{k})$ 
By definition, entropy determines the number of available states (or, classically, phase volume). Assuming that system spends comparable time in different available states we conclude that since the equilibrium must be the
most probable state it corresponds to the entropy maximum.
The statistical entropy is identified as the quantity (depending on $\rho$) which is maximized for a physical ensemble where the maximization of the entropy determines the physical distribution $\rho$. 
The  entropy in terms of a Boltzmann picture can be written:
\begin{equation}
S_\mathbb{B}= k_B\int_{\Gamma} - \rho \log \rho \frac{d\boldsymbol{p} d \boldsymbol{q}}{c} 
      -\lambda _1 \left( \int_{\Gamma} \frac{\boldsymbol{p}^2}{m^2}\rho \frac{d\boldsymbol{p} d \boldsymbol{q}}{c} -N\mu^2\right)
      -\lambda_2\left(\int_{\Gamma} \rho  \frac{d\boldsymbol{p} d \boldsymbol{q}}{c} -1  \right)
\end{equation}
where the second and the third constraint terms account for enforcing the root mean square velocity and for enforcing  normalization respectively.  Maximizing the entropy is equivalent to the  minimization of the information about the system. 

The maximization of entropy under constrains provide the value of the Lagrange multipliers $(\lambda_1,\lambda_2)$:
\begin{equation}
0=\delta S=\int_{\Gamma} \left(-\log\rho -1 -\frac{\lambda_1}{m^2}\boldsymbol{p}^2 -\lambda_2  \right)\delta \rho  \frac{d\boldsymbol{p} d \boldsymbol{q}}{c} 
\end{equation}

After some calculations we find that 
\begin{equation}
\rho = e^{-(1+\lambda_2)} e^{-\frac{\lambda_1\boldsymbol{p}^2}{m^2}}
\end{equation}
meanwhile from constraints we get the value of the multipliers:
\begin{align}
\int_{\Gamma} \frac{\boldsymbol{p}^2}{m^2}\rho \frac{d\boldsymbol{p} d \boldsymbol{q}}{c} = N\mu^2 & \Rightarrow \lambda_1=\frac{1}{\mu}\\
\int_{\Gamma} \rho  \frac{d\boldsymbol{p} d \boldsymbol{q}}{c} =1 & \Rightarrow  \left( \frac{m^2\pi }{\lambda_1}\right)^N \frac{1}{c} \int_{\Gamma} d\boldsymbol{q}=e^{1+\lambda_2}
\end{align}
The last integral $A_N=\int_{\Gamma} d\boldsymbol{q}$ represents the configurational spatial entropy which can be calculated as:
\begin{equation}
A_N= \frac{1}{N!}\prod_{n=1}^{N}(A-(n-1)4\pi r^2)
\end{equation}
where $A$ is the area of the region in which the population is considered, and $r$ is the radius of interaction of each identical individual which occupies a small region $\pi (2r)^2$.
Replacing and making some algebraic manipulation in the sparse limit we get the approximation of the uncorrelated Shanon entropy in terms of social distancing variables as:
\begin{equation}
S_\mathbb{U}= S_\mathbb{B}/(k_BN) \approx  2\log\frac{\mu}{\mu_1} +\log\frac{2\delta_1}{\delta} + 2
\end{equation}
where we do explicitly see the dependence of the entropy to all the social distancing and census data $(\mu,\delta| \mu_1,\delta_1)$ where $\delta_0=1/4\pi r^2$ interpreted as individual spacial density. Then, $\mu_1$ is the size of a geohash divided by the duration of a day, meaning that there is a minimal velocity of a device (or a group of) that cannot be detected because under the sensitivity of recording devices.  However the previous equation cannot be used directly as comparison to Shannon Entropy reported by data due to a lack of information and interpretation of the constants in the boltzamnn derivation of entropy.

Let us notice that in the possonian-type epidemic model where the infection rate is constant over time, the fundamental assumption relies on assuming the entropy constant over time, apart from some calibration, so that $r\sim 1/\sqrt{\mu}$ which is not in general true because of the unpredictability of the movement changes over time as consequences of social distancing behaviors.  
%

\subsection{Real entropy and Kolmogorov picture}\label{app_real}
Now we briefly discuss about the realistic effects of time-correlated patterns in human mobility measures.
Let us consider a set of locations in a target area, and a time-ordered input sequence $X = \{x_1 , x_2,\ldots, x_{n-1}\}$ where $x_i$ is the $i$-th location visited the user representing a random variable of the location at time $t_i, i = 1, 2,\ldots, n$. The duration at location $x_i$ is the time difference between $t_i$ and $t_{i+1}$.  In the study of human mobility we have introduce the concept of random entropy $S_{rand}$ measures the uncertainty of an individual's next location assuming that this individual's movement is completely random among $N$ possible locations. Then we have defined Shannon entropy $S_{\mathbb{U}}$  where the individual's movement among $N$ possible locations follows a given probability distribution and it is referred to as the temporal-uncorrelated entropy. There is another deinition of entropy which also takes in account the frequency of the visited locations and the order in which these locations are visited. It was introduce by \citet{song2010limits} and then used by many other authors \citet{wang2020entropy,kulkarni2019examining,teixeira2019deciphering,lu2013approaching}, it captures the temporal correlations among individual trajectories. It is defined as:
\begin{equation}
S_{real}=-\sum_{T'\in T}P(T')log(P(T'))
\end{equation}
where $T$ represents the sequence of the visited locations and $T'$ represents a subsequence of $T$.
However the real entropy  cannot be obtained directly using the previous relation but can be estimated by entropy rate estimators for example using Burrows-Wheeler transform (BWT) estimator or the Lempel-Ziv data compression algorithm \citet{kontoyiannis1998nonparametric,lempel1976complexity} as:
\begin{equation}
S_{real}\approx \frac{\log N}{\frac{1}{N}\sum_{n=1}^NL_i}
\end{equation} 
where $N$ is the length of the trajectory (total number of locations)  and $L_i$ is defined as the length of
the shortest substring at an index $i$ not appearing previously from index $1$ to $i - 1$. Both algorithms can be applied to observed transitions between locations and they are almost-sure convergent for stationary, ergodic random processes  characteristic of movement trajectories. In particular A person with a smaller real entropy is considered more predictable, as he is more constrained to the same sub-paths in the same order.  However, data we have used do not permit such fine analysis, and therefore we cannot extract any activity time related feature from our time patterns. Consequently, we cannot calculate directly the real entropy, but it is know that theoretically that $S_{real}\leq S_{\mathbb{U}}  \leq S_{rand}$. It is important to emphasize that when the process is completely random, $S_{real}= S_{\mathbb{U}}  = S_{rand}$, and when the process is not completely random but includes inherent repetitive patterns, the real entropy $S_{real}$ is the smallest among the three entropy rate measures. So, we expect to observe a real entropy which takes in account the changes on temporal pattern of individual movements which realistically are different from a no memory (posissonian) process.  Despite that fact it is important to notice that beeing able to reduce one entropy will always result in real entropy which will be never large than the shannon entropy. Furthermore one should also include cross-correlations in the path of different individuals, meaning that a trajectory of one individual can haxve an impact on the trajectory of another individual.  Despite such micro-founded approach to mobility, if one observe an aggregate mobility the Shannon entropy will result to be reliable measure of tracking the effectiveness of mobility control on the entropy and so on the controlling the unpredictability of human movements.

\section{Population Weighted Density}\label{app_pop}
\begin{figure}[!ht]
	\centering
	\begin{subfigure}[c]{0.75\textwidth}
		\centering
		\includegraphics[angle=0,origin=c,width=0.8\linewidth]{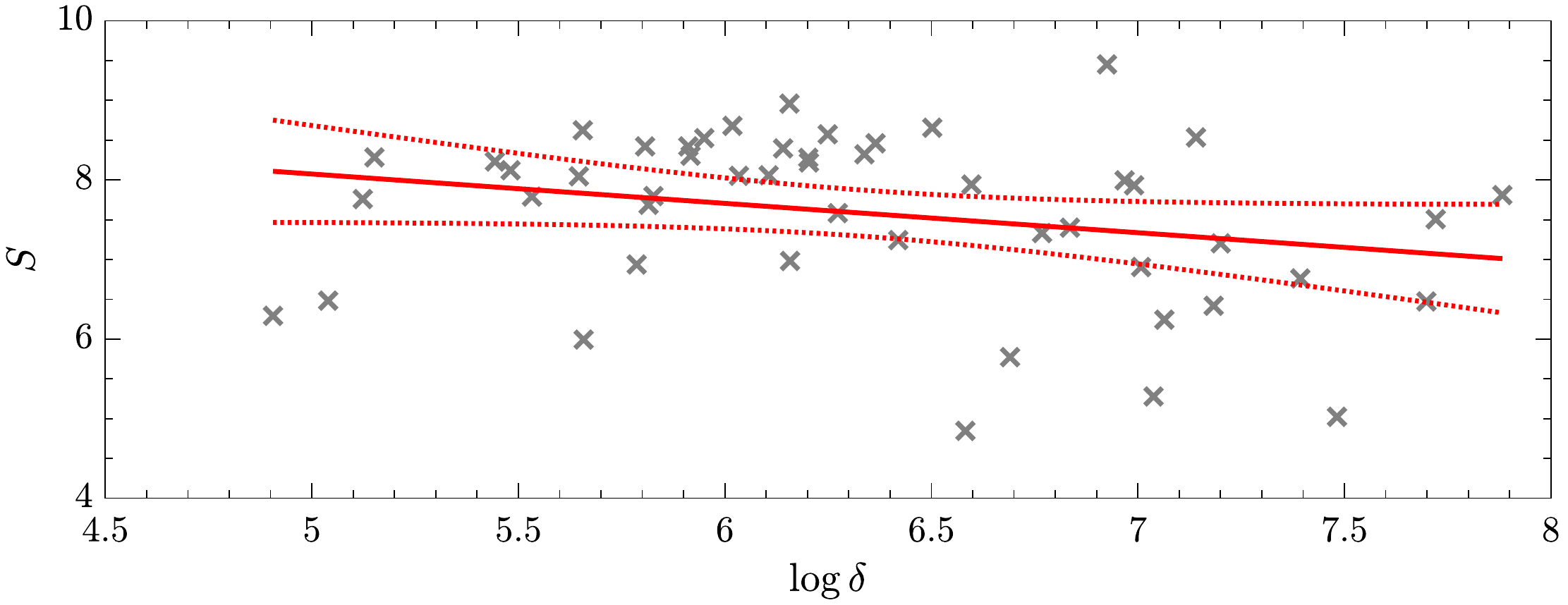}
		\caption{}
		\label{fig_densentropy}
	\end{subfigure}
	\\
	\vspace{0.5cm}
	\begin{subfigure}[c]{0.75\textwidth}
		\centering
		\includegraphics[angle=0,origin=c,width=0.8\linewidth]{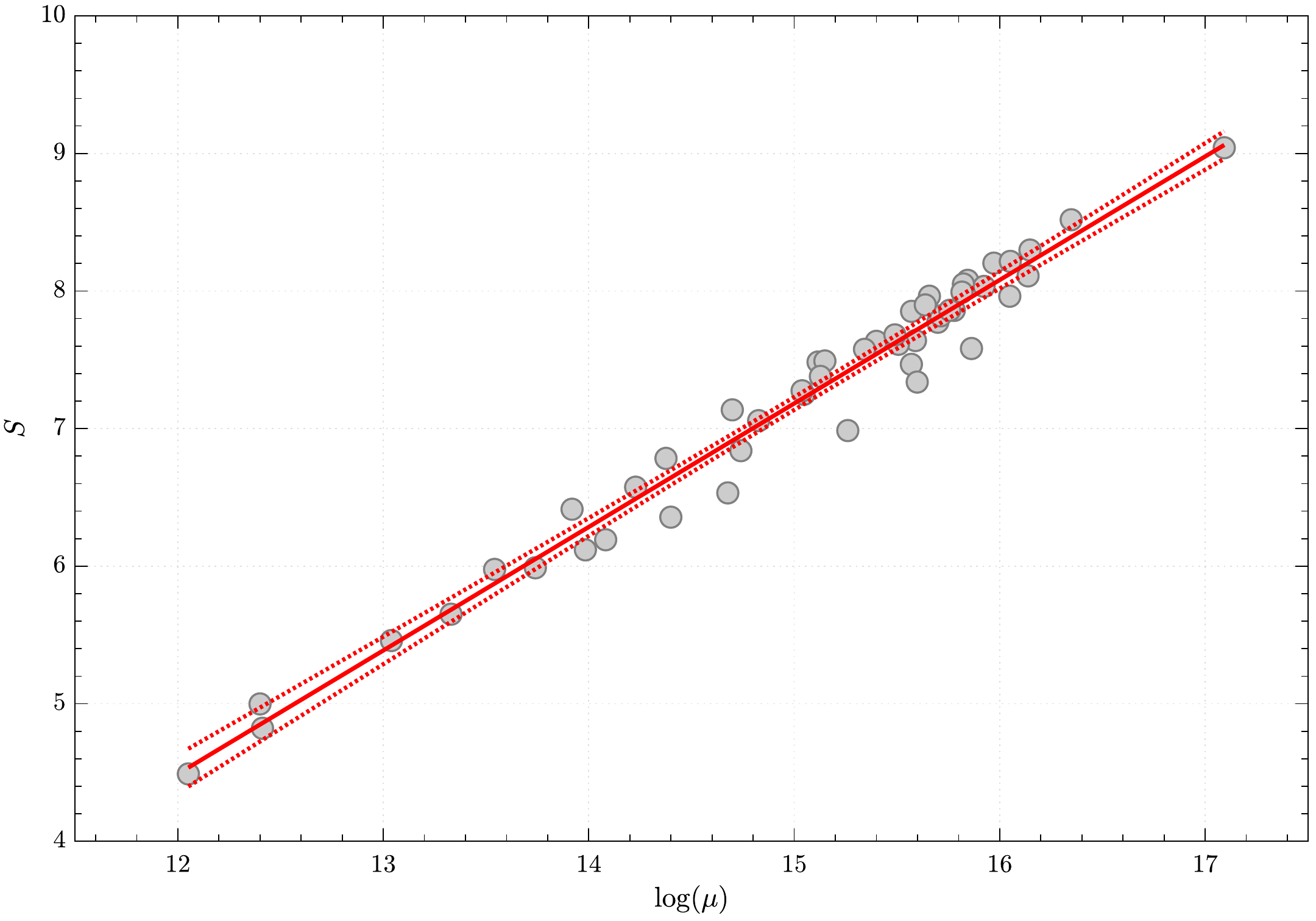}
		\caption{}\label{fig_mobentropy}
	\end{subfigure}
	\caption{Scatter plot analysis between the  random entropy and population weighted density (a) and human mobility  in terms of radius of gyration (b). The cross-sectional regression analysis has been performed for all the 50 states for a given period of time i.e. averaging of the first 14 days of February 2020.   Data courtesy of Camber Systems \citet{camber} for entropy and mobility measures and  WorldMap \citet{worldpopPWD1km} for population weighted density.}\label{fi_2dRegressionUS}
\end{figure}

We can observe in fig.\ref{fi_2dRegressionUS} as the entropy is related to two important social distancing variables as mobility and density among other factors which are among the main variables of interest in the interaction rate and the lockdown intensity in the epidemiological and economic functions respectively.
Such empirical results could be grounded by some theoretical models supported by statistical evidences which can be related with a collision framework of individual interactions responsible for the spreading of an epidemic.

However, density is one of the most fundamental properties of urban areas, what makes a city different from a suburb, and suburbs different from rural areas is chiefly how many people there are, and how close they are to each other.  The fact that people in cities live and work near each other is both economically and socially important.
 The ordinary gross population density is defined as the  population divided by area (more more specifically, land area) so depending on arbitrary boundaries.  However, it is a rather flawed measure since it deals with large geographic entities such as counties, states, and countries. Consequently this gross measure ignores the fact that most people live in structured regions as cities. As an economic measure, population density needs to accommodate the fact that most economic agents live much more concentrated in space than gross population density measures suggest.
The two basic issues with average population density calculations are the arbitrariness of defining borders and the fact that average population density focuses on the density of the average plot of land, not the density observed by the average person. In fact population-weighted density captures density as perceived by a randomly chosen individual. The population density for all census-blocks within a county and then computes population-weighted mean density. Population-weighted
density is meant to measure average ''experienced'' density and was popularized in economics \citet{glaeser,rappaport2008productivity,carozzi2020urban}.
Population density is a very important variable in many economic analyses and together with the organization of the city has played an essential role in the speed and the intensity of transmission of a epidemic and the effectiveness of shelter-in-place responses (staying home, avoiding travel).
Many research works have been looked at the relationship between density, mobility and interpersonal distances, with the productivity of different regions and urban development \citet{rappaport2008productivity}. 
There several ways to correct for the arbitrariness of defining borders, with the introduction of a sort of perceived population density. In particular, population weighted density  proposed by \citet{craig1984averaging} is a family of methods that weight the value of density by their corresponding population size in the aggregation process. In particular, to gain perspective on the densities at which people live  the population-weighted is derived from the densities of all the census tracts included within the boundary of the Core-based statistical area (CBSA) or Census Tract \citet{census2011, ottensmann2018population}.
For example, Covid-19 pandemic  has  led  to  a  reduced demand for housing in neighborhoods with high population density.
This can be mainly due to the  diminished  need  of  living  close  to  jobs  that are  telework-compatible  and  the  declining  value of  access  to  consumption  amenities, \citet{liu2020impact}. In fact,  the  pandemic's  negative  effect  on  the  demand  for  density  persisted  and  strengthened even after the first wave of the outbreak. 
The desirability and structure of cities are shaped by the strength of agglomeration and dispersion forces. But,  the  COVID-19  pandemic  has  re-introduced  the  danger  of  disease  transmission  as  a potentially  serious  dispersion  force  in  modern  cities. So that office workers no longer commute to crowded urban locations for work, and instead conduct businesses virtually at home, Consumption amenities such as restaurants have also seen much fewer visits, owing to policy mandates or consumers' concerns over the potential exposure to the virus in indoor public spaces 

In the case of the United States, we have taken population estimates for 2019 from the U.S. Bureau of Census and combined it with land area data (as provided by the 2010 census). For each U.S. county $i$ we have population $P_i$ and land area $A_i$.  So the weighted population density is calculated we have used Subnational level Population Weighted Density (PWD) metrics as produced using \citet{worldpopPWD1km}

\end{document}